\documentclass[letterpaper]{article} % DO NOT CHANGE THIS
\usepackage{aaai2026}  % DO NOT CHANGE THIS
\usepackage{times}  % DO NOT CHANGE THIS
\usepackage{helvet}  % DO NOT CHANGE THIS
\usepackage{courier}  % DO NOT CHANGE THIS
\usepackage[hyphens]{url}  % DO NOT CHANGE THIS
\usepackage{graphicx} % DO NOT CHANGE THIS
\def\UrlFont{\rm}  % DO NOT CHANGE THIS
\usepackage{natbib}  % DO NOT CHANGE THIS AND DO NOT ADD ANY OPTIONS TO IT
\usepackage{caption} % DO NOT CHANGE THIS AND DO NOT ADD ANY OPTIONS TO IT
\usepackage{algorithm}
\usepackage{algorithmic}

\usepackage{multirow, booktabs, xcolor}
\usepackage{amsmath}

\usepackage[most]{tcolorbox}

\usepackage{newfloat}
\usepackage{listings}
\DeclareCaptionStyle{ruled}{labelfont=normalfont,labelsep=colon,strut=off} % DO NOT CHANGE THIS
\floatstyle{ruled}
\newfloat{listing}{tb}{lst}{}
\floatname{listing}{Listing}
\title{Soft Filtering: Guiding Zero-shot Composed Image Retrieval \\with Prescriptive and Proscriptive Constraints}
\author {
    Youjin Jung\textsuperscript{\rm 1},
    Seongwoo Cho\textsuperscript{\rm 1},
    Hyun-seok Min\textsuperscript{\rm 2},
    Sungchul Choi\textsuperscript{\rm 1}
}
\affiliations {
    \textsuperscript{\rm 1}Major in Industrial Data Science \& Engineering, Department of Industrial and Data Engineering, Pukyong National University\\
    \textsuperscript{\rm 2}Tomocube Inc.\\
    wjd1dbwls@pukyong.ac.kr,
    jsw6872@pukyong.ac.kr,
    hsmin@tomocube.com,
    sc82@pknu.ac.kr
}

\usepackage{bibentry}
\begin{document}

\maketitle

\begin{abstract}
Composed Image Retrieval (CIR) aims to find a target image that aligns with user intent, expressed through a reference image and a modification text. While Zero-shot CIR (ZS-CIR) methods sidestep the need for labeled training data by leveraging pretrained vision-language models, they often rely on a single fused query that merges all descriptive cues of what the user wants—tending to dilute key information and failing to account for what they wish to avoid. Moreover, current CIR benchmarks assume a single correct target per query, overlooking the ambiguity in modification texts.
To address these challenges, we propose Soft Filtering with Textual constraints (SoFT), a training-free, plug-and-play filtering module for ZS-CIR. SoFT leverages multimodal large language models (LLMs) to extract two complementary constraints from the reference-modification pair: prescriptive (must-have) and proscriptive (must-avoid) constraints. These serve as semantic filters that reward or penalize candidate images to re-rank results, without modifying the base retrieval model or adding supervision.
In addition, we construct a two-stage dataset pipeline that refines CIR benchmarks. We first identify multiple plausible targets per query to construct multi-target triplets, capturing the open-ended nature of user intent. Then guide multimodal LLMs to rewrite the modification text to focus on one target, while referencing contrastive distractors to ensure precision. This enables more comprehensive and reliable evaluation under varying ambiguity levels.
Applied on top of CIReVL—a ZS-CIR retriever—SoFT raises $R@5$ to 65.25 on CIRR (+12.94), $mAP@50$ to 27.93 on CIRCO (+6.13), and $R@50$ to 58.44 on FashionIQ (+4.59), demonstrating broad effectiveness.
\end{abstract}

\begin{links}
    \link{Code}{https://github.com/jjungyujin/SoFT}
    \link{Datasets}{https://github.com/jjungyujin/SoFT/blob/main/MultiTarget_README.md}
\end{links}

\section{Introduction}
\begin{figure}[t]
    \centering
    \includegraphics[width=1.0\linewidth]{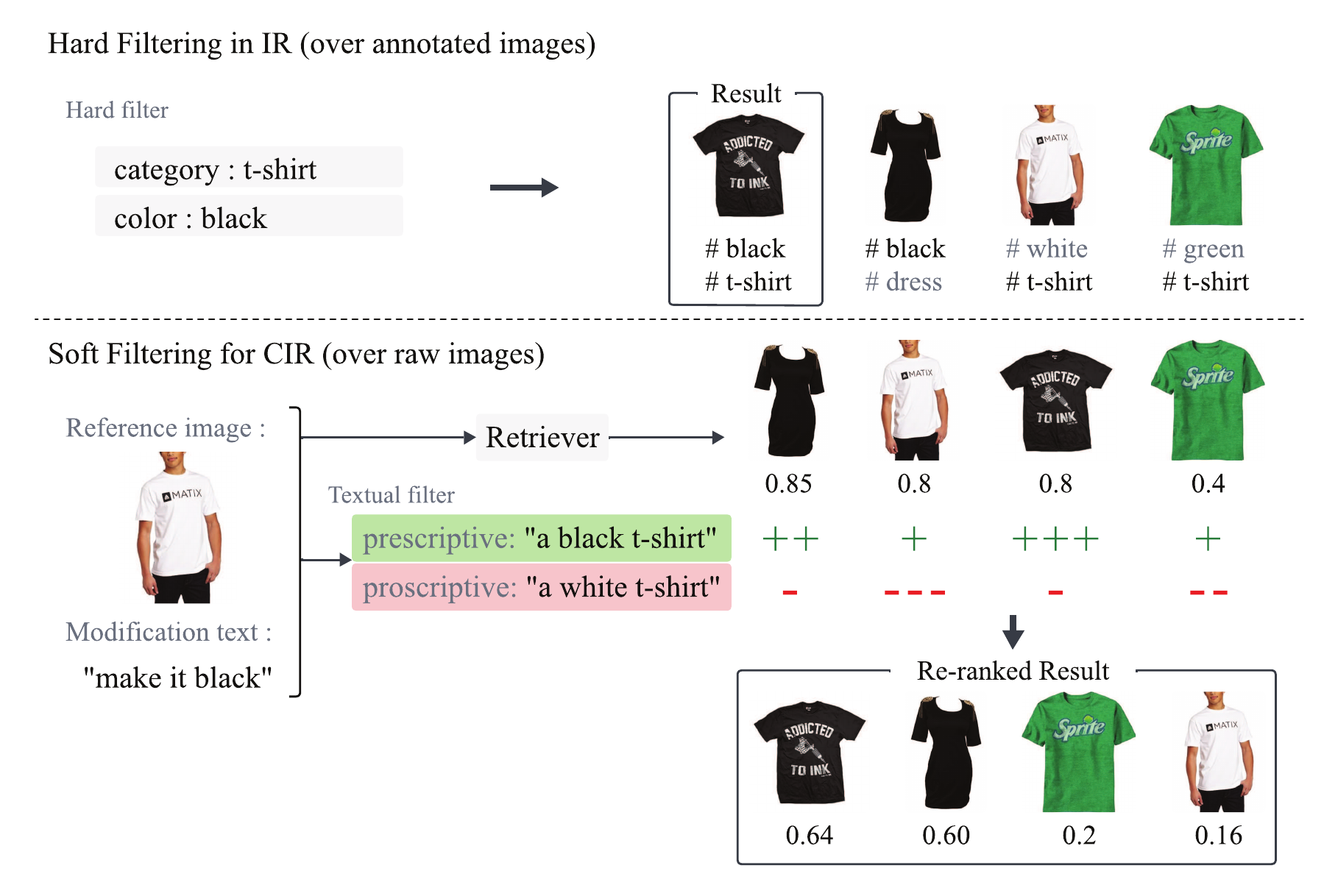}
    \caption{Comparison of hard filtering in traditional IR and soft filtering for CIR. SoFT re-ranks unstructured image candidates using LLM-generated constraints.}
    \label{fig:hard_soft_filtering}
\end{figure}

Information retrieval (IR)~\cite{schutze2008introduction} enables users to find relevant content based on natural language queries. With recent advances in multi-modal representation learning, the field has extended into visual domains, enabling more expressive and personalized search paradigms. One such task is Composed Image Retrieval (CIR)~\cite{vo2019composing, lee2021cosmo, combiner, chen2020image, hosseinzadeh2020composed}, where the goal is to retrieve a target image based on a reference image and a natural language modification that describes the desired change. However, the construction of triplet datasets—each consisting of a reference image, a modification text, and a matching target image—for supervised CIR~\cite{delmasartemis, dodds2020modality, liu2021image} is costly and often domain-specific, limiting scalability. To address this issue, zero-shot CIR (ZS-CIR) has emerged, aiming to eliminate the need for task-specific training and improve generalization to novel compositions~\cite{pic2word, searle, gu2024language}.

While recent ZS-CIR methods~\cite{collm, cirevl, osrcir} have made progress by leveraging pretrained vision-language models~\cite{clip, song2022clip, zhou2022conditional, jia2021scaling}, most still rely on single-query matching strategies. In these approaches, all descriptive cues—both visual and textual—are compressed into a single representation, regardless of their relative importance. As a result, critical user requirements that should be strongly enforced are diluted by less relevant visual or textual details, compromising retrieval accuracy~\cite{ldre}. Moreover, such methods often overlook the need to penalize undesired attributes in retrieval, focusing only on satisfying positive cues. LDRE~\cite{ldre} addresses this gap by proposing an ensemble-based approach. However, it still does not explicitly disentangle and control the prescriptive (must-have) and proscriptive (must-avoid) aspects of user intent, as it operates over variations of fused representations.

To address these limitations, we propose \textbf{Soft Filtering with Textual constraints (SoFT)}, filtering mechanism tailored for the CIR. Unlike traditional retrieval systems that apply hard filters over structured metadata~\cite{niu2019understanding, yee2003faceted, zhao2017memory}, CIR operates over raw image-text pairs without explicit annotations, rendering conventional filtering inapplicable. Inspired by classical IR systems that enforce conditions to reduce irrelevant results, SoFT adapts this concept to the unstructured, multimodal nature of CIR, as illustrated in Figure~\ref{fig:hard_soft_filtering}.

Instead of hard exclusion, SoFT re-ranks candidate images by interpreting user intent through \textbf{dual-faceted textual constraints}. We leverage multimodal Large Language Models (LLMs) to automatically generate two complementary constraints from the reference-modification pair: a \textbf{prescriptive} constraint emphasizing attributes that the target should include, and a \textbf{proscriptive} constraint describing attributes that should be avoided. These guide similarity-based reward and penalty scores for each candidate image, enabling precise, constraint-aware re-ranking—without any additional training or annotations.

Moreover, current ZS-CIR evaluation benchmarks~\cite{fiq, cirr} overlook the ambiguities in modification texts, assuming a single correct target per query despite the existence of multiple valid answers~\cite{fiq, collm}. To address this, we introduce a \textbf{multi-target triplet construction pipeline}, which first identifies semantically valid targets per query, then rewrite the modification text via LLMs to generate single-target triplets. This process could also be used to enrich existing multi-target datasets with diversified constraints.

Our contributions are twofold:
\begin{itemize}
\item We propose SoFT, a training-free re-ranking module that leverages prescriptive and proscriptive textual constraints to improve retrieval precision.
\item We develop a two-stage dataset pipeline that captures both ambiguity and specificity in user intent by constructing multi-target triplets and deriving single-target variants through LLM-guided refinement.
\end{itemize}

\section{Related Works}
\subsubsection{Zero-shot Composed Image Retrieval and LLM-based Reasoning.}
CIR aims to retrieve target images based on a reference image and a modification text describing the desired change. Traditional CIR~\cite{vo2019composing, chen2020image, chen2020learning, shin2021rtic, lee2021cosmo, delmas2022artemis, baldrati2022effective} methods rely on supervised learning over carefully curated triplet datasets (reference image, modification text, and target image), which are labor-intensive and costly to construct. To solve this problem, ZS-CIR approaches have emerged, eliminating the need for triplet supervision and enabling retrieval using pretrained models~\cite{pic2word, searle}.
Most ZS-CIR methods adopt pretrained CLIP~\cite{clip} as the retrieval backbone due to its strong vision-language alignment and zero-shot generalization capabilities. Early methods can be broadly categorized into fusion-based and inversion-based approaches. Fusion-based models~\cite{combiner, collm} encode the reference image and modification text independently before merging their representations. Inversion-based models~\cite{pic2word, searle, gu2024language, tang2024context} reinterpret the image as a pseudo-text token and perform joint encoding with the modification text. While both approaches avoid triplet training, they still require training dedicated modules for fusion or inversion.

Recently, LLMs have been adopted to enhance compositional reasoning in a fully training-free manner. For instance, CIReVL~\cite{cirevl} reformulates CIR as a two-step text generation process: first generating a caption for the reference image, then producing a modified query conditioned on the reference caption and the modification text. This formulation avoids rigid template matching (e.g., “a photo of X that Y”) and improves generalization to novel compositions.
OSrCIR~\cite{osrcir} simplifies the pipeline further by simultaneously providing both the reference image and modification text to the LLM, which directly generates a query tailored to the target image. This direct generation reduces the risk of information loss caused by intermediate representations and allows for more flexible reasoning.
LDRE~\cite{ldre} leverages LLMs to produce multiple pseudo queries, aggregating their retrieval scores for more robust performance. However, these methods still fall short in explicitly modeling the dual nature of user intent—what must be included and what must be avoided.

\begin{figure*}[t]
    \centering
    \includegraphics[width=1.0\linewidth]{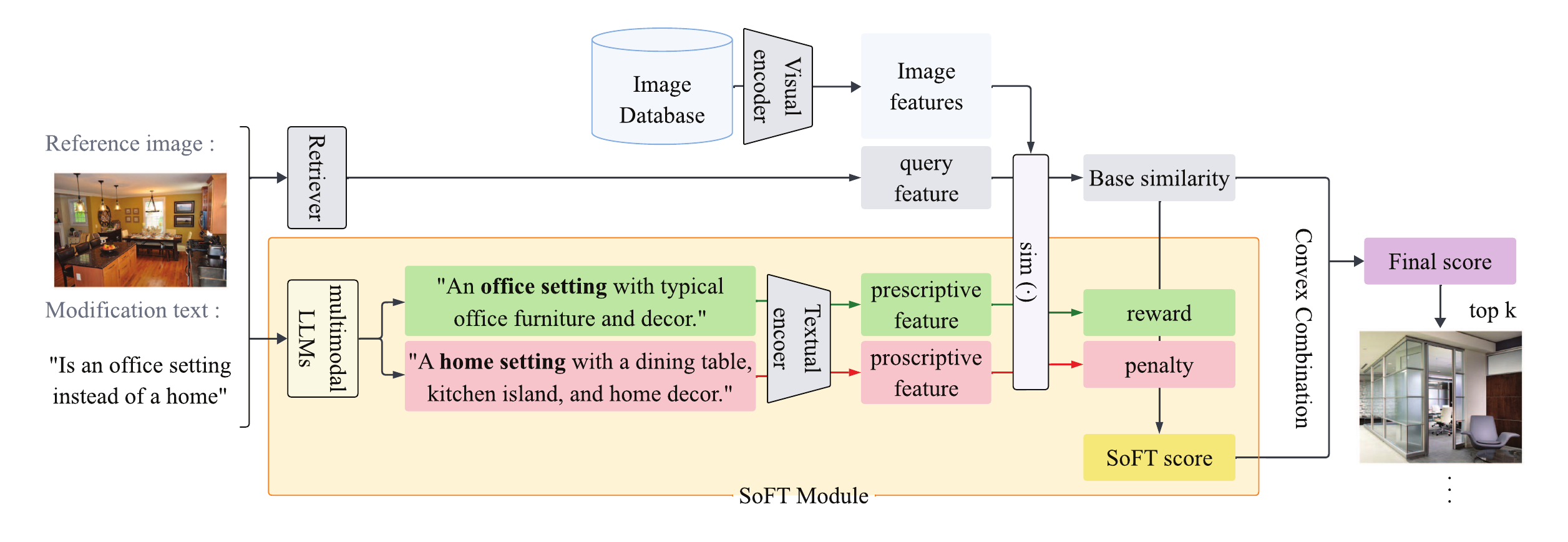}
    \caption{Overview of SoFT, a plug-and-play soft filtering module for Zero-shot CIR. Given a reference image and a modification text, multimodal LLMs extract prescriptive and proscriptive constraints. These are used to softly reward or penalize candidate images using CLIP similarity.}
    \label{fig:soft_main}
\end{figure*}

\subsubsection{CIR Benchmark Datasets.}
CIR has been primarily evaluated on datasets such as CIRCO~\cite{circo}, CIRR~\cite{cirr} and FashionIQ~\cite{fiq}. CIRCO introduces diverse and inherently ambiguous retrieval scenarios, CIRR emphasizes natural language composition in generic scenes, and FashionIQ focuses on fashion-related attribute modification. However, a common limitation in CIRR and FashionIQ is the assumption of a single correct target image, which fails to reflect the inherent ambiguity and multi-target validity that can arise from modification texts. These texts often lack specificity, leading to query ambiguity that impairs evaluation reliability.

CoLLM~\cite{collm} mitigates the ambiguity of modification texts by rewriting them via LLMs into more target-specific and discriminative instructions, reducing the likelihood of multiple plausible targets per query. While CoLLM improves retrieval evaluation by clarifying ambiguous modification texts, it still operates under the assumption of a single correct target. In contrast, our proposed dataset construction pipeline explicitly acknowledges the existence of multiple semantically valid targets, and leverages LLMs not only to identify diverse ground-truth candidates but also to generate disambiguated single-target queries, thereby enabling both multi-target and fine-grained single-target evaluations within a unified framework.

\section{Method}
We propose a two-fold approach to enhance ZS-CIR: (1) a soft filtering module based on dual textual constraints that operates without additional training, and (2) a multi-target dataset pipeline that enables more diverse and precise evaluation of retrieval models.

\subsection{Soft Filtering with Dual Textual Constraints}
CIR takes a reference image and a modification text describing the desired transformation as input. Even without explicit negations, the task naturally defines both \textbf{prescriptive} (must-have) and \textbf{proscriptive} (must-avoid) constraints. For instance, if the modification is “make it black” and the reference shows a white T-shirt (as in Figure~\ref{fig:hard_soft_filtering}), then the white color becomes an attribute to avoid in the target image.
We exploit this duality by decomposing the intent and reweighting candidate image scores accordingly, in a plug-and-play manner, as illustrated in the overview shown in Figure~\ref{fig:soft_main}.

\subsubsection{Dual Constraint Generation via LLM Prompting.}
To extract constraints from reference image \(I_\mathrm{ref}\) and modification text \(T_\mathrm{mod}\), we prompt an instruction-tuned LLM in two step. Full prompt template is provided in the Appendix.

In Step 1: Attribute Classification, the LLM identifies key attribute-value pairs and categorizes them into three types: \textit{keep} attributes, which should be retained from the reference image; \textit{add} attributes, which are newly required according to the modification; and \textit{remove} attributes, which are present in the reference but should be eliminated in the target.
Unlike generic descriptions, the LLM is instructed to infer attributes critical for achieving the transformation.

In Step 2: Constraint Generation, these sets are converted into:
\begin{itemize}
    \item a \textit{prescriptive constraint},(combining \textit{keep} and \textit{add}), and
    \item a \textit{proscriptive constraint} (based on \textit{remove}).
\end{itemize}
These serve as semantic filters, enforcing bidirectional control over the retrieval output.

\subsubsection{Soft Reweighting with Dual Constraints.}
Given the two constraints, we adjust candidate similarity score without modifying the underlying model.
For each candidate image \(I_c\), we compute CLIP-based cosine similarities in the joint embedding space:
\begin{itemize}
    \item \(s_\mathrm{base}\): similarity between \((I_\mathrm{ref}, T_\mathrm{mod})\) and \(I_c\) from a CIR model;
    \item \(s_\mathrm{reward}\): similarity between \(I_c\) and the prescriptive constraint;
    \item \(s_\mathrm{penalty}\): similarity between \(I_c\) and the proscriptive constraint.
\end{itemize}

We then compute a SoFT score \(s_\mathrm{SoFT}\) defined as:
\begin{equation}
    s_\mathrm{SoFT} = s_\mathrm{base} \odot \frac{s_\mathrm{reward} + 1 - s_\mathrm{penalty}}{2}
\label{eq:soft_score}
\end{equation}

Finally, the reweighted similarity score \(s_\mathrm{final}\) used for ranking is given by:
\begin{equation}
    s_\mathrm{final} = (1 - \lambda)s_\mathrm{base} + \lambda s_\mathrm{SoFT}
\label{eq:final_score}
\end{equation}

\begin{figure*}[t]
    \centering
    \includegraphics[width=0.95\linewidth]{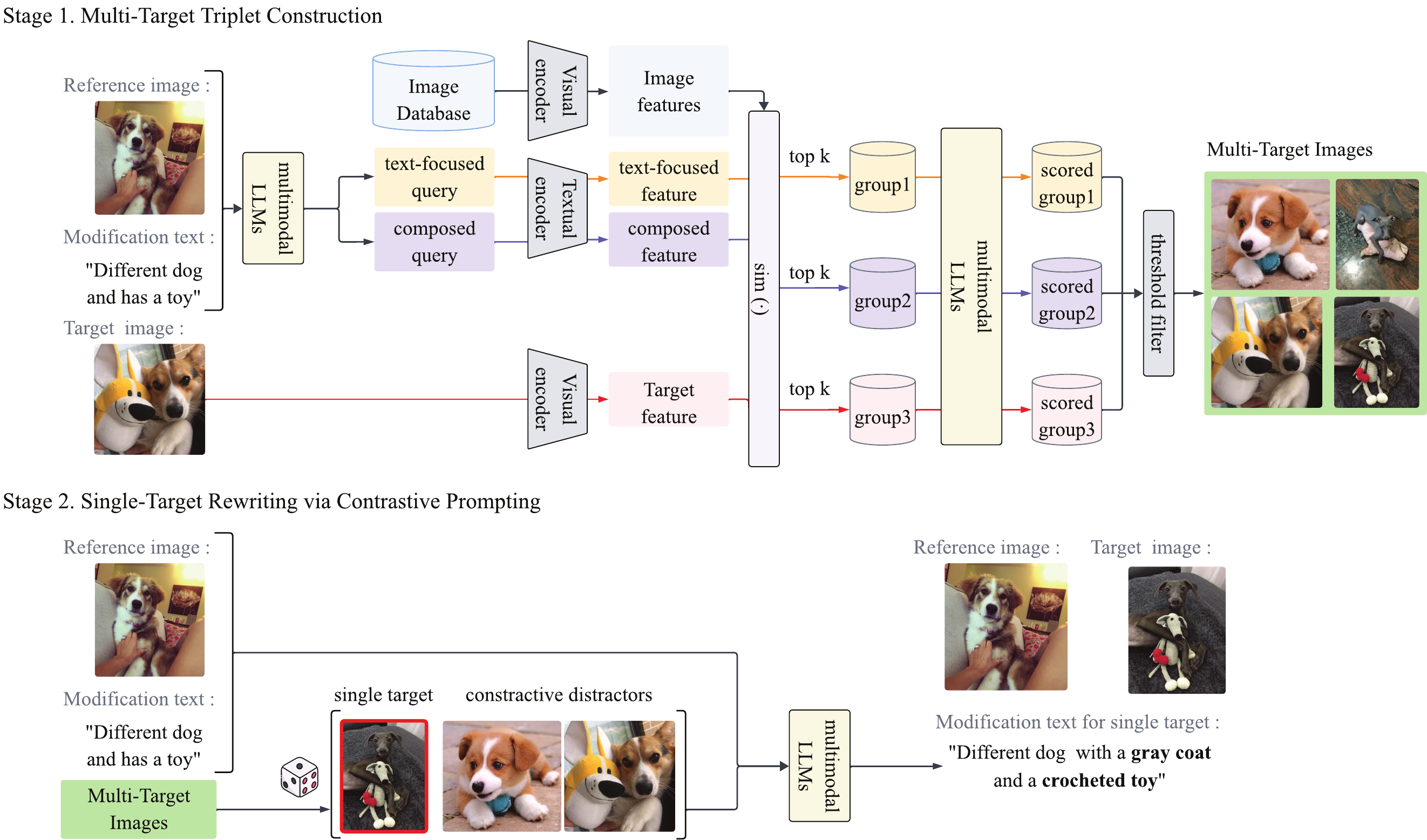}
    \caption{Overview of the Multi-target Triplet Dataset pipeline. Stage 1 selects diverse target images for each reference-modification pair to capture open-ended user intent. Stage 2 rewrites the modification text to focus on a specific target with contrastive distractors, enabling precise single-target evaluation.}
    \label{fig:dataset_pipeline}
\end{figure*}

This formulation enables \textbf{soft constraint enforcement} by promoting candidates aligned with the prescriptive intent while down-weighting those associated with the proscriptive constraint. The final score, computed as a convex combination of the base and filtered scores, remains continuous, interpretable, and tunable.

Importantly, SoFT is \textbf{model-agnostic and training-free}, operating entirely at inference time via score-level modulation. Its \textbf{plug-and-play} nature allows seamless integration into any CIR systems, making it especially suitable for zero-shot settings where architectural changes or supervision are impractical.

\subsection{Multi-Target Triplet Dataset Pipeline}
Most CIR benchmarks assume a single ground-truth image per query, which limits evaluation under ambiguous or open-ended user intents. To address this, we propose a two-stage pipeline that expands CIR datasets with multi-target triplets and refines them into discriminative single-target triplets. Stage 1 identifies multiple valid targets using both visual and textual similarity signals. Stage 2 rewrites the modification text to distinguish one target from semantically similar distractors. This design supports both inclusive and precise evaluation. An overview is illustrated in Figure~\ref{fig:dataset_pipeline}.

\subsubsection{Stage 1: Multi-Target Triplet Construction.}
We first retrieve a diverse set of plausible targets for each (reference image, modification text) pair using three CLIP-based criteria:
\begin{itemize}
    \item Textual to Modification: Top-k images similar to a text query describing the modification only.
    \item Compositional: Top-k images similar to a composed query combining the modification and non-conflicting details from the reference image.
    \item Visual Similarity to Original Target: Top-k images similar to the original ground-truth image.
\end{itemize}
The textual queries in (1) and (2) are generated by prompting an LLM to produce two concise sentences: one for the modification, and another preserving non-conflicting aspects of the reference. This allows fine-grained, compositional retrieval.

From the three groups of top-k candidates—each corresponding to one retrieval criterion—we apply an LLM-based visual assessor to assign a confidence score between 0.0 and 1.0 for each image. Scoring is performed group-wise, with the reference image and modification text provided as context. Any image that receives a score above a predefined threshold (denoted \(\tau\)) in its respective group is selected as a valid multi-target. We use \(\tau = 0.85\) and \(k = 10\) by default. This approach ensures high-quality semantic coverage while requiring no manual annotation.

\begin{table*}[t]
\centering
\setlength{\tabcolsep}{3pt}
\renewcommand{\arraystretch}{1.2}
\begin{tabular}{ll|cccc|cccc|ccc}
\toprule
\multicolumn{2}{c|}{\textbf{Benchmark}} & \multicolumn{4}{c|}{\textbf{CIRCO (mAP@K)}} & \multicolumn{4}{c|}{\textbf{CIRR (Recall@K)}} & \multicolumn{3}{c}{\textbf{CIRR (Recall$_\mathrm{subset}$@K)}} \\
\textbf{Backbone} & \textbf{Method} & k=5 & k=10 & k=25 & k=50 & k=1 & k=5 & k=10 & k=50 & k=1 & k=2 & k=3 \\
\midrule
\multirow{6}{*}{ViT-B/32}
& SEARLE & 9.35 & 9.94 & 11.13 & 11.84 & 24.00 & 53.42 & 66.82 & 89.78 & 54.89 & 76.60 & 88.19 \\
& SEARLE + SoFT & 12.57 & 13.12 & 14.38 & 15.17 & 28.05 & 58.60 & 71.88 & 91.78 & 64.29 & 82.19 & 91.61 \\
& CIReVL & 14.94 & 15.42 & 17.00 & 17.82 & 23.94 & 52.51 & 66.00 & 86.95 & 60.17 & 80.05 & 90.19 \\
& \textbf{CIReVL + SoFT} & \textbf{19.21} & \textbf{20.04} & \textbf{21.86} & \textbf{22.75} & \textbf{32.94} & \textbf{62.92} & \textbf{74.17} & \textbf{90.70} & \textbf{70.31} & \textbf{86.58} & \textbf{93.78} \\
& LDRE & 17.96 & 18.32 & 20.21 & 21.11 & 25.69 & 55.13 & 69.04 & 89.90 & 60.53 & 80.65 & 90.70 \\
& OSrCIR & 18.04 & 19.17 & 20.94 & 21.85 & 25.42 & 54.54 & 68.19 & N/A & 62.31 & 80.86 & 91.13 \\

\midrule
\multirow{6}{*}{ViT-L/14}
& SEARLE & 11.68 & 12.73 & 14.33 & 15.12 & 24.24 & 52.48 & 66.29 & 88.33 & 53.76 & 75.01 & 88.19 \\
& SEARLE + SoFT & 15.72 & 16.45 & 18.06 & 18.93 & 30.29 & 59.74 & 71.49 & 90.65 & 65.47 & 83.23 & 92.29 \\
& CIReVL & 18.57 & 19.01 & 20.89 & 21.80 & 24.55 & 52.31 & 64.92 & 86.34 & 59.54 & 79.88 & 89.69 \\
& \textbf{CIReVL + SoFT} & \textbf{23.90} & 24.72 & 26.94 & 27.93 & \textbf{35.54} & \textbf{65.25} & \textbf{76.41} & \textbf{91.95} & \textbf{71.59} & \textbf{87.64} & \textbf{94.15} \\
& LDRE & 23.35 & 24.03 & 26.44 & 27.50 & 26.53 & 55.57 & 67.54 & 88.50 & 60.43 & 80.31 & 89.90 \\
& OSrCIR & 23.87 & \textbf{25.33} & \textbf{27.84} & \textbf{28.97} & 29.45 & 57.68 & 69.86 & N/A & 62.12 & 81.92 & 91.10 \\
\bottomrule
\end{tabular}
\caption{Quantitative results on CIRCO and CIRR.}
\label{tab:soft-circo-cirr}
\end{table*}

\subsubsection{Stage 2: Single-Target Rewriting via Contrastive Prompting.}
To evaluate fine-grained discrimination, we convert multi-target pools into single-target triplets. From each pool, we randomly sample one target and two contrastive distractors. An LLM is then prompted with the reference image, original modification, target, and distractors to rewrite the modification text. The new text must preserve the reference context, uniquely describe the target, and exclude distractors. This yields more challenging triplets that test precise, compositional reasoning—complementing the broader evaluation enabled by multi-targets.

\begin{figure}[t]
    \centering
    \includegraphics[width=\linewidth]{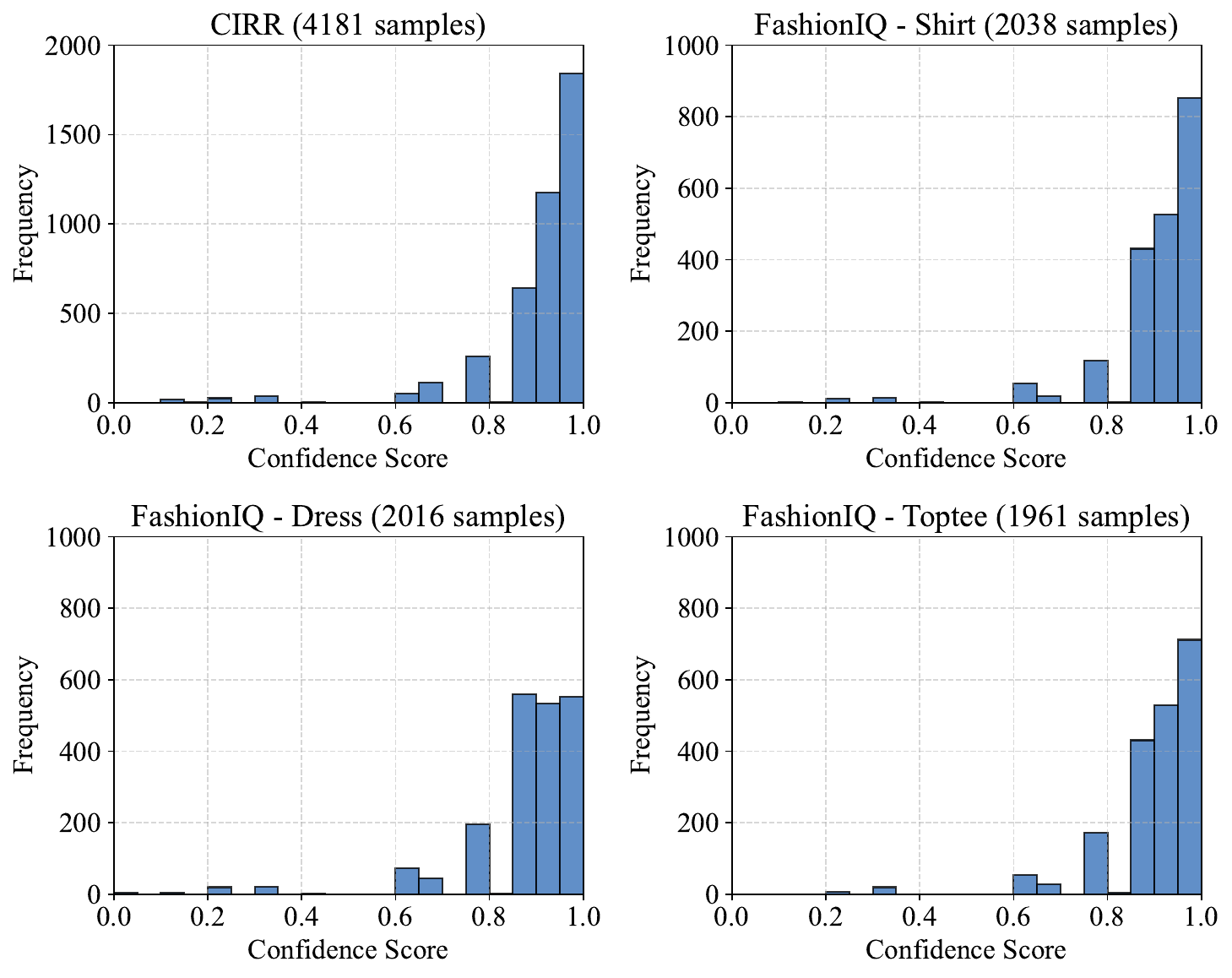}
    \caption{
        Confidence scores assigned by the LLM-based evaluator to original ground-truth target images across each dataset, illustrating the consistency and reliability of the scoring method.
    }
\label{fig:target_confidence_distribution}
\end{figure}

\subsubsection{Dataset Application and Statistics.}
We apply our pipeline to the validation splits of and CIRR and FashionIQ. Queries with no valid alternative target is identified—i.e., all candidates fall below the confidence threshold—are excluded, slightly reducing in dataset size: CIRR decreases from 4,181 to 4,140 queries, while FashionIQ retains 2,032 for Shirt, 2,011 for Dress and 1,958 for Toptee. Our method substantially enriches each query with multiple valid targets, yielding an average of 2.89 for CIRR, 4.68 (Shirt), 4.93 (Dress) and 4.60 (Toptee) targets for FashionIQ. This diverse target pool supports more reliable evaluation under open-ended user intent. Figure~\ref{fig:target_confidence_distribution} shows the confidence scores assigned to original benchmark targets by our LLM-based scoring module.

Building on these multi-target sets, we further construct single-target triplets to support precise evaluation under contrastive conditions. 
This stage is applied only to queries with at least three valid targets from Stage 1, ensuring that one target and two contrastive distractors can be selected from the same candidate pool. 
As a result, single-target triplets are constructed for 2,245 CIRR queries, and for 1,681, 1,827 and 1,666 queries in the Shirt, Dress and Toptee subsets of FashionIQ, respectively. 
As illustrated in Figure~\ref{fig:dataset_pipeline}, the refined texts introduce precise descriptors—e.g., “gray coat”, “crocheted toy”—that improve discriminative power in high-similarity scenarios. 
More qualitative examples of the constructed triplets for both CIRR and FashionIQ are provided in the Appendix.

\section{Experiments}
\subsection{Experimental Setup}
\subsubsection{Datasets.}
We evaluate SoFT on three standard CIR benchmarks-CIRCO~\cite{circo}, CIRR~\cite{cirr}, and FashionIQ~\cite{fiq}-as well as our extended Multi-Target variants. CIRCO is constructed from the COCO 2017 unlabeled set~\cite{lin2014microsoft}, provides multiple valid targets per query. CIRR contains real-world scene images and is designed to test fine-grained reasoning via retrieval within visually similar subsets. FashionIQ is a fashion-focused benchmark comprising triplets across Shirt, Dress and Toptee categories, with evaluation conducted on the validation split. For both CIRCO and CIRR, evaluation is performed on the official test split by submitting prediction files to the respective evaluation servers.

\subsubsection{Baselines.}
We compare our method against two publicly available CIR baselines: SEARLE~\cite{searle} and CIReVL~\cite{cirevl}. Both provide full implementations and serve as reliable references for evaluating plug-and-play compatibility. Reference results from recent models such as LDRE~\cite{ldre} and OSrCIR~\cite{osrcir} are also included in tables for context.

\subsubsection{Evaluation Metrics.}
We report Recall (\(R@K\)) for single-target benchmarks (CIRR, FashionIQ), measuring the presence of the ground-truth image within the top-K results. For multi-target datasets (CIRCO and our Multi-Target variants), mean Average Precision (\(mAP@k\)) is used, which rewards ranking lists that place valid targets higher and more consistently.

\subsubsection{Implementation Details.}
For all components of our framework, including both the proposed SoFT module and the dataset construction pipeline, we consistently use GPT-4o~\cite{hurst2024gpt} via OpenAI’s API with temperature set to 0.0. All similarity computations are based on pretrained CLIP models~\cite{clip}, using the inner product of encoder outputs. In SoFT, the final retrieval score is computed as in Eq.~\eqref{eq:final_score}, with \(\lambda=1.0\) for CIReVL and \(\lambda=0.2\) for SEARLE. Main experiments compare ViT-B/32 and ViT-L/14~\cite{dosovitskiy2020image}; the latter is used by default elsewhere. 

\subsubsection{Computation and Cost Analysis.}
All experiments were conducted on a single NVIDIA RTX 4090 GPU. 
CLIP-based feature extraction and similarity computations were performed locally on GPU for efficiency, 
while all SoFT and dataset pipeline operations involving language models were executed via the GPT-4o API.
The LLM cost for SoFT was \$2.69 for CIRCO, \$15.36 for CIRR, and \$17.68 for FashionIQ. For dataset construction, the total cost was \$196.40 (Stage 1) and \$12.31 (Stage 2) on CIRR, and \$116.32 (Stage 1) and \$16.38 (Stage 2) on FashionIQ.

\begin{table}[t]
    \centering
    {\fontsize{9}{11}\selectfont
    \setlength{\tabcolsep}{0.7mm}
    \begin{tabular}{l|cc|cc|cc|cc}
    \toprule
     & \multicolumn{2}{c|}{\textbf{Shirt}} & \multicolumn{2}{c|}{\textbf{Dress}} & \multicolumn{2}{c|}{\textbf{TopTee}} & \multicolumn{2}{c}{\textbf{Average}} \\
    \textbf{Recall@k} & @10 & @50 & @10 & @50 & @10 & @50 & @10 & @50 \\
    \midrule
    \multicolumn{6}{l}{\textbf{Method}} \\
    \multicolumn{6}{l}{\textbf{ViT-B/32}} \\
    SEARLE & 24.44 & 41.61 & 18.54 & 39.51 & 25.70 & 46.46 & 22.89 & 42.53 \\
    SEARLE$^*$ & 24.29 & 42.64 & 19.48 & 40.46 & 25.54 & 47.78 & 23.10 & 43.63 \\
    CIReVL & 28.36&	47.84	&25.29&	46.36	&31.21	&53.85&	28.29	&49.35 \\
    CIReVL$^*$ &31.26&	50.98	&27.52&	49.13	&34.37	&58.44&	31.05
	&52.85
    \\
    LDRE & 27.38 & 46.27 & 19.97 & 41.84 & 27.07 & 48.78 & 24.81 & 45.63 \\
    OSrCIR & 31.16 & 51.13 & 29.35 & 50.37 & 36.51 & 58.71 & 32.34 & 53.40 \\
    \midrule
    \multicolumn{6}{l}{\textbf{ViT-L/14}} \\
    SEARLE & 26.89 & 45.58 & 20.48 & 43.13 & 29.32 & 49.97 & 25.56 & 46.23 \\
    SEARLE$^*$ & 26.30 & 43.67 & 20.77 & 42.94 & 27.94 & 49.97 & 25.00 & 45.53 \\
    CIReVL & 29.49&	47.40	&24.79	&44.76&	31.36&	53.65	&28.55	&48.57 \\
    CIReVL$^*$ & 32.34&	51.62	&27.51
&	47.79	&35.19	&58.18	&31.68	&52.53 \\
    LDRE & 31.04 & 51.22 & 22.93 & 46.76 & 31.57 & 53.64 & 28.51 & 50.54 \\
    OSrCIR & 33.17 & 52.03 & 29.70 & 51.81 & 36.92 & 59.27 & 33.26 & 54.37 \\
    \bottomrule
    \end{tabular}
    }
    \caption{Quantitative results on FashionIQ. $^*$ denotes methods with our SoFT module.}
    \label{tab:soft-fiq}
\end{table}

\subsection{Effectiveness of Soft Filtering Module}
As shown in Table~\ref{tab:soft-circo-cirr}, SoFT consistently improves retrieval performance across all metrics and backbones on CIRCO and CIRR. On CIRCO with ViT-B/32, it enhances \(mAP@5\) from 9.35 to 12.57 (+3.22) and \(mAP@50\) from 11.84 to 15.17 (+3.33) for SEARLE. For CIReVL, the improvements are more substantial: \(mAP@5\) increases from 14.94 to 19.21 (+4.27) and \(mAP@50\) from from 17.82 to 22.75 (+4.93). With the stronger ViT-L/14 backbone, CIReVL show the most significant gains—mAP@5 rises from 18.57 to 23.90 (+5.33), and mAP@50 from 21.80 to 27.93 (+6.13). On CIRR, SoFT also yields consistent gains, especially in the \textit{subset} split, which demands finer-grained disambiguation. For example, when applied to SEARLE with ViT-B/32, \(R@1\) on the subset improves from 54.89 to 64.29, surpassing both LDRE (60.53) and OSrCIR (62.31). CIReVL+SoFT further raises \(R@1\) from 60.17 to 70.31 (+10.14), again outperforming all baselines. Similar trends are observed with ViT-L/14.

Table~\ref{tab:soft-fiq} shows that SoFT also benefits CIReVL on the FashionIQ, with average improvements of +2.76 (R@10) and +3.5 (R@50) on ViT-B/32, and even higher gains on ViT-L/14. In contrast, SEARLE exhibits mixed results, showing improvements in some cases but slight declines in others. These fluctuations may stem from the differing reasoning mechanisms of the two retrievers.
Nevertheless, it is noteworthy that on Multi-Target FashionIQ, SoFT consistently improves performance regardless of the base model (see Sec~\ref{sec:mt-fiq-cirr}). This suggests that while baseline characteristics affect stability, the occasional drops on the standard FashionIQ are more likely due to the dataset’s limitation—implicitly overlooking semantically valid alternative targets rather than reflecting a failure of SoFT itself.

To ensure that the improvements stem from the SoFT rather than LLM differences, we unified CIReVL (originally using GPT-3.5-turbo) and SoFT under GPT-4o and repeated the experiments. The results, reported in the Appendix, show consistent gains across all three benchmarks.

\begin{table}[t]
    \centering
    {\fontsize{9}{11}\selectfont
    \setlength{\tabcolsep}{0.7mm}
    \begin{tabular}{l|cc|cc|cc|cc}
    \toprule
     & \multicolumn{2}{c|}{\textbf{Shirt}} & \multicolumn{2}{c|}{\textbf{Dress}} & \multicolumn{2}{c|}{\textbf{TopTee}} & \multicolumn{2}{c}{\textbf{Average}} \\
    \textbf{mAP@k} & @5 & @25 & @5 & @25 & @5 & @25 & @5 & @25 \\
    \midrule
    \multicolumn{6}{l}{\textbf{Method}} \\
    \multicolumn{6}{l}{\textbf{ViT-L/14}} \\
    CIReVL & 24.14 & 23.36 & 20.71 & 19.62 & 23.85 & 23.17 & 22.90 & 22.05 \\
    CIReVL$^*$ & 28.17 & 26.67 & 24.39 & 22.31 & 28.86 & 27.05 & 27.14 & 27.04 \\
    SEARLE & 41.64 & 36.54 & 36.26 & 31.46 & 44.44 & 38.53 & 40.78 & 35.51 \\
    SEARLE$^*$ & 42.45 & 37.21 & 37.50 & 32.44 & 45.37 & 39.14 & 41.77 & 36.26 \\
    \bottomrule
    \end{tabular}
    }
    \caption{Retrieval performance(mAP@5/mAP@25) on the Multi-Target FashionIQ validation set. $^*$ indicates methods with our SoFT module applied. SoFT consistently improves mAP@5 and mAP@25 across all categories and base models.}
    \label{tab:soft-mt-fiq}
\end{table}

\subsection{Re-evaluation on Multi-Target FashionIQ.}
\label{sec:mt-fiq-cirr}
We re-evaluate the SoFT on our multi-target version of the FashionIQ validation set. As shown in Table~\ref{tab:soft-mt-fiq}, both SEARLE and CIReVL consistently benefit from SoFT across all categories, improving R@10 and R@50. We additionally investigate the effect of the convex combination weight \(\lambda\), observing that SEARLE achieves its best performance when \(\lambda = 1\), with peak scores of 45.50 (\(mAP@5\)) and 39.05 (\(mAP@25\)) on average. Additional experiments, including Multi-Target CIRR results and examples of dataset, are provided in the Appendix.

\begin{figure}[t]
    \centering
    \includegraphics[width=\linewidth]{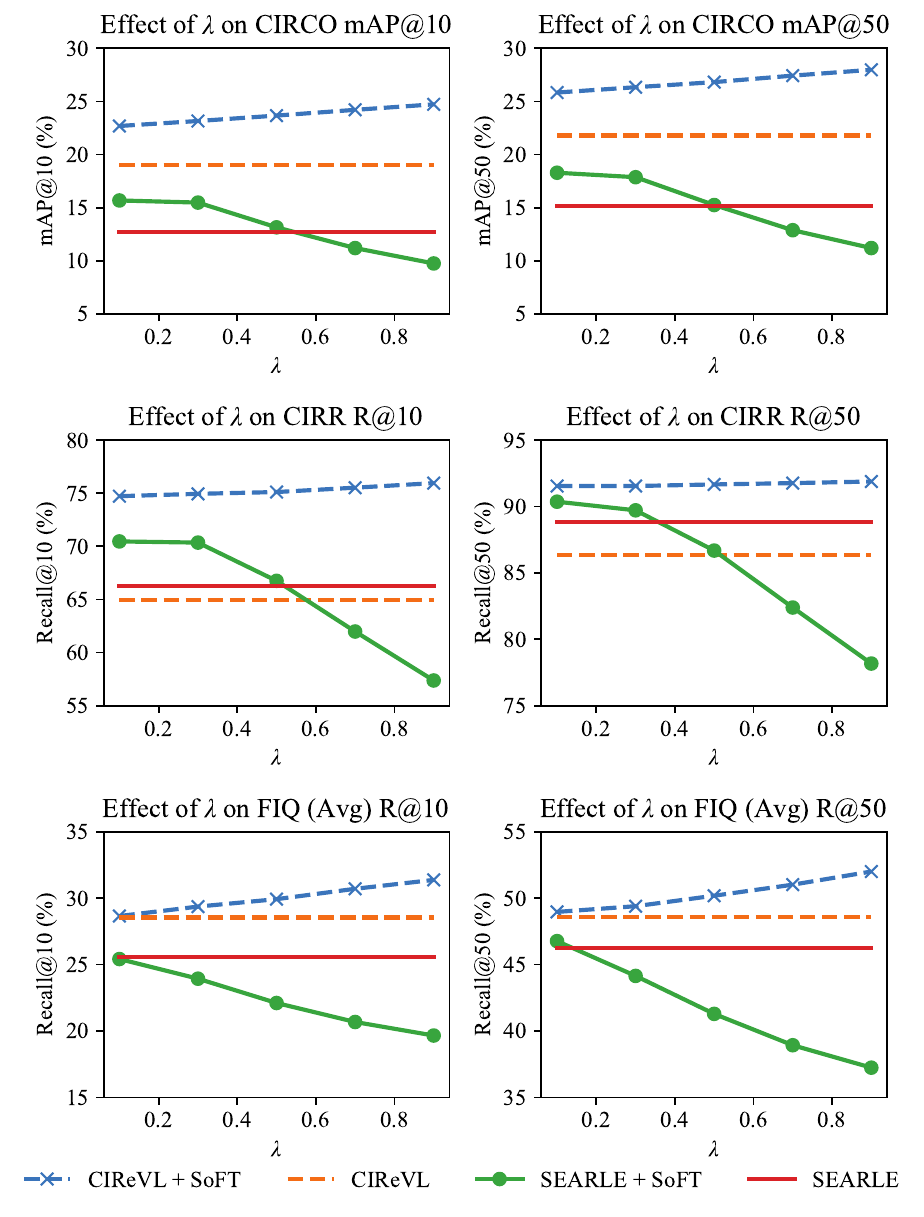}
    \caption{Effect of the weighting parameter \(\lambda\) on retrieval performance across CIRCO, CIRR, and FashionIQ (CLIP L/14). We vary \(\lambda \in \{0.1, 0.3, 0.5, 0.7, 0.9\}\), controlling the influence of SoFT in Equation~\eqref{eq:final_score}.}
    \label{fig:ablation_by_lambda}
\end{figure}

\subsection{Ablation Studies}
\subsubsection{Effect of the weight \(\lambda\).} 
We examine how the interpolation weight $\lambda$ in Equation~\eqref{eq:final_score} affects performance, varying $\lambda \in \{0.1, 0.3, 0.5, 0.7, 0.9\}$ on CIRCO, CIRR, and FashionIQ. As shown in Figure~\ref{fig:ablation_by_lambda}, SoFT consistently improves CIReVL across all benchmarks, indicating that constraint-guided reweighting complements the base similarity effectively. In contrast, SEARLE exhibits sensitivity to the choice of $\lambda$, with performance generally decreasing as $\lambda$ increases across benchmarks. Together with the contrasting trends observed for different baselines, these results suggest that the balance between the base similarity and SoFT’s filtering signals may vary depending on the retriever. Consequently, treating $\lambda$ as a tunable parameter rather than a fixed constant may be beneficial for achieving stable performance. Full results are provided in Appendix.

\begin{table}[t]
\centering
    {\fontsize{9}{11}\selectfont
    \setlength{\tabcolsep}{1mm}
    \begin{tabular}{l|cc|cc|cc}
    \toprule
     & \multicolumn{2}{c|}{\textbf{CIRCO}} & \multicolumn{2}{c|}{\textbf{CIRR}} & \multicolumn{2}{c}{\textbf{FIQ (Avg)}} \\
      & \multicolumn{2}{c|}{mAP@k} & \multicolumn{2}{c|}{Recall@k} & \multicolumn{2}{c}{Recall@k} \\
    \textbf{Method} & k=5 & k=50 & k=1 & k=10 & k=10 & k=50 \\
    \midrule
    SEARLE & 11.68 & 15.12 & 24.24 & 66.29 & 25.56 & 46.23 \\
    SEARLE + Penalty & 0.28 & 0.28 & 4.82 & 15.49 & 3.59 & 7.63 \\
    SEARLE + Reward & 18.06 & 22.15 & 29.08 & 70.84 & 25.59 & 46.42 \\
    SEARLE + SoFT & 15.72 & 18.93 & 30.29 & 71.49 & 25.00 & 45.53 \\
    CIReVL & 18.57 & 21.80 & 24.55 & 64.92 & 28.55 & 48.57 \\
    CIReVL + Penalty & 21.83 & 25.57 & 32.92 & 74.51 & 28.77 & 47.13 \\
    CIReVL + Reward & 22.19 & 26.00 & 32.80 & 74.99 & 32.57 & 53.46 \\
    CIReVL + SoFT & 23.90 & 27.93 & 35.54 & 76.41 & 31.68 & 52.53 \\
    \bottomrule
    \end{tabular}
    }
\caption{Component-wise ablation of SoFT on CIRCO, CIRR, and FashionIQ. We compare three variants: applying only the reward term (\texttt{+Reward}, $s_\mathrm{SoFT} = s_\mathrm{base} \cdot s_\mathrm{reward}$), only the penalty term (\texttt{+Penalty}, $s_\mathrm{SoFT} = s_\mathrm{base} \cdot (1 - s_\mathrm{penalty})$), and both (\texttt{+SoFT}, as defined in Equation~\eqref{eq:soft_score}).}
\label{tab:ablation_by_component}
\end{table}

\subsubsection{Component-wise Analysis of SoFT.}
To ablate the contributions of each constraint in SoFT by comparing three variants: using only the reward term (\texttt{+Reward}), only the penalty term (\texttt{+Penalty}), and both (\texttt{+SoFT}), as shown in Table~\ref{tab:ablation_by_component}. On CIReVL, both constraints yield additive gains, with the full SoFT achieving the best performance. In contrast, SEARLE shows strong sensitivity to the penalty term: using it alone leads to a severe performance drop. Nevertheless, when combined with the reward term, SoFT consistently outperforms the baseline, underscoring the importance of jointly maintaining prescriptive and proscriptive constraints for balanced filtering.
In contrast, on FashionIQ, instability induced by the penalty term is observed regardless of the underlying retriever. This behavior can be attributed to the limitations of pretrained CLIP representations: in fine-grained domains, CLIP embeddings often fail to capture subtle visual distinctions, which may result in over-penalizing semantically valid candidates. As a consequence, the penalty signal can become unreliable when applied in isolation, highlighting the need for balanced integration of proscriptive constraints.

\section{Conclusion}
We propose SoFT, a training-free, plug-and-play filtering module for Zero-shot Composed Image Retrieval (ZS-CIR) that leverages dual textual constraints—prescriptive and proscriptive—to re-rank retrieval candidates. SoFT addresses the limitations of single fused queries by explicitly capturing both positive and negative aspects of user intent.
We also introduce a two-stage dataset pipeline that expands CIR benchmarks with multi-target triplets and contrastively refined single-target variants. This better reflects the ambiguity and diversity of user intent in real-world queries. Applied on top of existing retrievers such as CIReVL and SEARLE, SoFT consistently improves retrieval performance across CIRCO, CIRR and FashionIQ, as well as on the newly constructed Multi-Target variants of CIRR and FashionIQ. These results demonstrate robust generalization and stable gains without any additional training or parameter tuning.

\clearpage

\section{Acknowledgments}
This work was supported by the IITP (Institute of Information \& Communications Technology Planning \& Evaluation)-ICAN (ICT Challenge and Advanced Network of HRD) (IITP-2024-RS-2023-00259806, 20\%) grant funded by the Korea government(Ministry of Science and ICT) and the National Research Foundation of Korea (NRF) grant funded by the Korea government (MSIT) (RS-2024-00354675, 40\%) and (RS-2024-00352184, 40\%).

\bibliography{aaai2026}

\clearpage
\makeatletter
\setcounter{table}{0}
\renewcommand{\thetable}{S\arabic{table}}    

\setcounter{figure}{0}
\renewcommand{\thefigure}{S\arabic{figure}}

\appendix
\section*{Appendix}
\section{Soft Filtering with Dual Textual Constraints}
\subsection{Large Language Model Prompt Templates}
The SoFT module utilizes a unified prompt, termed the Dual Constraint Extraction Prompt, to extract both prescriptive (must-have) and proscriptive (must-avoid) textual constraints from each reference-modification pair. This prompt is designed to be dataset-agnostic and generalizable across domains. Accordingly, it is applied consistently in all experiments, including those on CIRCO, CIRR, FashionIQ and our constructed multi-target benchmarks.

\begin{tcolorbox}[colback=red!5,colframe=red!40!black,title=Dual Constraint Extraction Prompt Template]
You are a helpful assistant for a composed image retrieval system.
You are given a reference image and a modification text that describes how the image should be changed.
Your task is to extract meaningful attribute information and generate two types of semantic queries from a reference image and a modification text. \\

\#\# Step 1. Attribute Classification \\
Analyze the modification text in the context of the reference image and extract three types of attribute-value lists. \\
- ``keep'': A list of key attribute values that should remain unchanged in the reference image according to the modification text. \\
- ``add'': A list of attribute values that are not present in the reference image but are explicitly or implicitly required by the modification text. \\
- ``remove'': A list of attribute values that are present in the reference image but are explicitly removed by the modification text. \\

\#\# Step 2. Query Generation \\
Using the attribute-value lists from Step 1, generate text queries that can be directly used for image retrieval.
Each query must be fluent, self-contained. If an attribute is described in relative terms, you must resolve it into a concrete absolute caption using visual clues from the reference image. \\
- ``prescriptive\_query'': Write a short, specific image caption that describes the visual features that should be included in the target image, focusing on the attribute values in the ``keep'' and ``add'' lists. \\
- ``proscriptive\_query'': Write a short, specific image caption that describes the reference image as it is, using positive, natural language, focusing on the attribute values in the ``remove'' list. \\

\#\# Input: \\
Modification Text: ``\{mod\_text\}'' \\
Reference image: \\
\end{tcolorbox}

\begin{table}
    \centering
    {\setlength{\tabcolsep}{1.8mm}
        \begin{tabular}{l|cccc}
        \toprule
        Method & \multicolumn{4}{c}{\textbf{CIRCO (mAP@k)}} \\
        & @5 & @10 & @25 & @50 \\
        \midrule
        \multicolumn{5}{l}{\textbf{ViT-B/32}} \\
        SEARLE & 9.35 & 9.94 & 11.13 & 11.84 \\
        SEARLE + SoFT (0.1) & 11.47 & 12.14 & 13.39 & 14.14 \\
        SEARLE + SoFT (0.2) & \textbf{12.57} & \textbf{13.12} & \textbf{14.38} & \textbf{15.17} \\
        SEARLE + SoFT (0.3) & 12.37 & 12.89 & 14.25 & 15.00 \\
        SEARLE + SoFT (0.5) & 11.33 & 11.61 & 12.86 & 13.59 \\
        SEARLE + SoFT (0.7) & 9.49 & 9.72 & 10.81 & 11.40 \\
        SEARLE + SoFT (0.9) & 8.19 & 8.38 & 9.23 & 9.72 \\
        CIReVL & 14.94 & 15.42 & 17.00 & 17.82 \\
        CIReVL + SoFT (0.1) & 17.68 & 18.32 & 20.02 & 20.91 \\
        CIReVL + SoFT (0.3) & 17.98 & 18.61 & 20.35 & 21.27 \\
        CIReVL + SoFT (0.5) & 18.33 & 18.99 & 20.72 & 21.64 \\
        CIReVL + SoFT (0.7) & 18.65 & 19.37 & 21.13 & 22.08 \\
        CIReVL + SoFT (0.9) & 19.16 & 19.81 & 21.65 & 22.55 \\
        CIReVL + SoFT (1.0) & \textbf{19.21} & \textbf{20.04} & \textbf{21.86} & \textbf{22.75} \\
        \midrule
        \multicolumn{5}{l}{\textbf{ViT-L/14}} \\
        SEARLE & 11.68 & 12.73 & 14.33 & 15.12 \\
        SEARLE + SoFT (0.1) & 15.07 & 15.68 & 17.43 & 18.29 \\
        SEARLE + SoFT (0.2) & \textbf{15.72} & \textbf{16.45} & \textbf{18.06} & \textbf{18.93} \\
        SEARLE + SoFT (0.3) & 14.76 & 15.48 & 17.03 & 17.87 \\
        SEARLE + SoFT (0.5) & 12.42 & 13.15 & 14.51 & 15.25 \\
        SEARLE + SoFT (0.7) & 10.72 & 11.20 & 12.25 & 12.88 \\
        SEARLE + SoFT (0.9) & 9.27 & 9.75 & 10.67 & 11.20 \\
        CIReVL & 18.57 & 19.01 & 20.89 & 21.80 \\
        CIReVL + SoFT (0.1) & 22.10 & 22.69 & 24.88 & 25.84 \\
        CIReVL + SoFT (0.3) & 22.45 & 23.16 & 25.37 & 26.34 \\
        CIReVL + SoFT (0.5) & 22.90 & 23.67 & 25.83 & 26.82 \\
        CIReVL + SoFT (0.7) & 23.48 & 24.21 & 26.42 & 27.43 \\
        CIReVL + SoFT (0.9) & \textbf{23.97} & \textbf{24.73} & \textbf{26.99} & \textbf{27.98} \\
        CIReVL + SoFT (1.0) & 23.90 & 24.72 & 26.94 & 27.93 \\
        \bottomrule
        \end{tabular}
    }
    \caption{Effect of interpolation weight $\lambda$ on CIRCO retrieval performance.}
    \label{tab:lambda-circo}
\end{table}

\subsection{Quantitative Results}
Tables~\ref{tab:lambda-circo}, \ref{tab:lambda-cirr}, and \ref{tab:lambda-fiq} show the effect of interpolation weight $\lambda$ in the following convex combination of scores:
\[
s_{final} = (1 - \lambda) \cdot s_{\text{base}} + \lambda \cdot s_{\text{SoFT}},
\]
where $s_{\text{base}}$ is the original retrieval score and $s_{\text{SoFT}}$ is the adjustment derived from the reward and penalty constraints. SEARLE demonstrates the most stable performance when $\lambda = 0.2$, while CIReVL improves steadily up to $\lambda = 1.0$. Accordingly, we set $\lambda = 0.2$ for SEARLE and $\lambda = 1.0$ for CIReVL as the default values in subsequent analyses.

Tables~\ref{tab:component-circo}, \ref{tab:component-cirr}, and \ref{tab:component-fiq} present a component-wise analysis of SoFT, where each constraint—Reward and Penalty—is applied individually or in combination. 
Overall, the combination of both constraints generally leads to the highest scores, indicating that their synergy tends to enhance retrieval precision and robustness. At the same time, applying only the reward constraint yields more stable performance than using the penalty constraint alone, suggesting that positive guidance aligns more consistently with the base retrieval signal.

\begin{table*}
    \centering
    {
        \begin{tabular}{l|cccc|ccc}
        \toprule
        Method & \multicolumn{4}{c|}{\textbf{CIRR}} & \multicolumn{3}{c}{\textbf{CIRR Subset}} \\
        & R@1 & R@5 & R@10 & R@50 & R@1 & R@2 & R@3 \\
        \midrule
        \multicolumn{8}{l}{\textbf{ViT-B/32}} \\
        SEARLE & 24.00 & 53.42 & 66.82 & 89.78 & 54.89 & 76.60 & 88.19 \\
        SEARLE + SoFT (0.1) & 27.16 & 57.42 & 70.27 & 91.16 & 60.02 & 79.81 & 90.02 \\
        SEARLE + SoFT (0.2) & 28.05 & \textbf{58.60} & \textbf{71.88} & \textbf{91.78} & 64.29 & 82.19 & 91.61 \\
        SEARLE + SoFT (0.3) & \textbf{28.46} & 58.27 & 71.35 & 91.35 & 66.68 & 83.86 & 92.43 \\
        SEARLE + SoFT (0.5) & 27.45 & 55.74 & 68.48 & 88.77 & 68.48 & \textbf{85.42} & 93.30 \\
        SEARLE + SoFT (0.7) & 25.25 & 52.22 & 63.81 & 85.25 & \textbf{68.51} & \textbf{85.42} & \textbf{93.61} \\
        SEARLE + SoFT (0.9) & 22.51 & 47.40 & 58.39 & 80.68 & 67.78 & 85.04 & 93.28 \\
        CIReVL & 23.94 & 52.51 & 66.00 & 86.95 & 60.17 & 80.05 & 90.19 \\
        CIReVL + SoFT (0.1) & 30.46 & 61.13 & 72.22 & 90.29 & 65.90 & 83.83 & 92.41 \\
        CIReVL + SoFT (0.3) & 31.25 & 61.49 & 72.60 & 90.41 & 67.06 & 84.24 & 92.70 \\
        CIReVL + SoFT (0.5) & 31.71 & 61.66 & 72.99 & 90.60 & 67.95 & 84.60 & 92.77 \\
        CIReVL + SoFT (0.7) & 32.22 & 62.19 & 73.25 & \textbf{90.70} & 68.77 & 85.18 & 93.13 \\
        CIReVL + SoFT (0.9) & 32.77 & 62.58 & 73.90 & \textbf{90.70} & 69.64 & 85.85 & 93.37 \\
        CIReVL + SoFT (1.0) & \textbf{32.94} & \textbf{62.92} & \textbf{74.17} & \textbf{90.70} & \textbf{70.31} & \textbf{86.58} & \textbf{93.78} \\
        \midrule
        \multicolumn{8}{l}{\textbf{ViT-L/14}} \\
        SEARLE & 24.24 & 52.48 & 66.29 & 88.84 & 53.76 & 75.01 & 88.19 \\
        SEARLE + SoFT (0.1) & 28.65 & 58.80 & 70.46 & 90.36 & 61.25 & 80.43 & 90.31 \\
        SEARLE + SoFT (0.2) & \textbf{30.29} & \textbf{59.74} & \textbf{71.49} & \textbf{90.65} & 65.47 & 83.23 & 92.29 \\
        SEARLE + SoFT (0.3) & 30.19 & 58.70 & 70.35 & 89.71 & 67.35 & 85.11 & \textbf{92.99} \\
        SEARLE + SoFT (0.5) & 27.71 & 55.30 & 66.75 & 86.68 & 68.22 & \textbf{85.71} & 92.96 \\
        SEARLE + SoFT (0.7) & 25.78 & 50.96 & 61.98 & 82.39 & \textbf{68.29} & 85.49 & 92.75 \\
        SEARLE + SoFT (0.9) & 23.76 & 47.35 & 57.37 & 78.17 & 67.74 & 84.48 & 92.07 \\
        CIReVL & 24.55 & 52.31 & 64.92 & 86.34 & 59.54 & 79.88 & 89.69 \\
        CIReVL + SoFT (0.1) & 32.87 & 63.28 & 74.72 & 91.54 & 67.35 & 85.98 & 93.49 \\
        CIReVL + SoFT (0.3) & 33.54 & 63.71 & 74.94 & 91.54 & 67.88 & 86.12 & 93.61 \\
        CIReVL + SoFT (0.5) & 33.64 & 64.12 & 75.11 & 91.66 & 68.75 & 86.36 & 93.81 \\
        CIReVL + SoFT (0.7) & 34.36 & 64.55 & 75.52 & 91.76 & 69.78 & 87.08 & 93.86 \\
        CIReVL + SoFT (0.9) & 34.96 & 64.72 & 75.95 & 91.88 & 70.65 & 87.47 & 94.07 \\
        CIReVL + SoFT (1.0) & \textbf{35.54} & \textbf{65.25} & \textbf{76.41} & \textbf{91.95} & \textbf{71.59} & \textbf{87.64} & \textbf{94.15} \\
        \bottomrule
        \end{tabular}
    }
    \caption{Effect of interpolation weight $\lambda$ on CIRR retrieval performance.}
    \label{tab:lambda-cirr}
\end{table*}

\begin{table*}
    \centering
    {\setlength{\tabcolsep}{1.9mm}
    \begin{tabular}{l|cc|cc|cc|cc}
        \toprule
        Method & \multicolumn{2}{c|}{\textbf{Shirt}} & \multicolumn{2}{c|}{\textbf{Dress}} & \multicolumn{2}{c|}{\textbf{TopTee}} & \multicolumn{2}{c}{\textbf{Average}} \\
        & R@10 & R@50 & R@10 & R@50 & R@10 & R@50 & R@10 & R@50 \\
        \midrule
        \multicolumn{9}{l}{\textbf{ViT-B/32}} \\
        SEARLE & \textbf{24.44} & 41.61 & 18.54 & 39.51 & 25.70 & 46.46 & 22.89 & 42.53 \\
        SEARLE + SoFT (0.1) & 24.43 & \textbf{42.69} & 19.43 & 40.01 & \textbf{26.52} & \textbf{48.09} & \textbf{23.46} & 43.60 \\
        SEARLE + SoFT (0.2) & 24.29 & 42.64 & \textbf{19.48} & \textbf{40.46} & 25.54 & 47.78 & 23.10 & \textbf{43.63} \\
        SEARLE + SoFT (0.3) & 23.65 & 41.27 & 19.29 & 39.46 & 24.89 & 46.30 & 22.61 & 42.34 \\
        SEARLE + SoFT (0.5) & 21.64 & 38.86 & 17.90 & 36.79 & 22.39 & 43.24 & 20.64 & 39.63 \\
        SEARLE + SoFT (0.7) & 19.87 & 36.21 & 16.71 & 34.80 & 20.75 & 40.74 & 19.11 & 37.25 \\
        SEARLE + SoFT (0.9) & 18.20 & 34.25 & 15.47 & 33.42 & 19.43 & 37.53 & 17.70 & 35.07 \\
        CIReVL & 28.36 & 47.84 & 25.29 & 46.36 & 31.21 & 53.85 & 28.29 & 49.35 \\
        CIReVL + SoFT (0.1) & 28.56 & 47.55 & 25.14 & 46.11 & 30.80 & 54.36 & 28.17 & 49.34 \\
        CIReVL + SoFT (0.3) & 29.44 & 47.74 & 25.68 & 46.26 & 31.77 & 54.87 & 28.96 & 49.62 \\
        CIReVL + SoFT (0.5) & 29.69 & 48.28 & 26.08 & 46.95 & 32.74 & 55.84 & 29.50 & 50.36 \\
        CIReVL + SoFT (0.7) & 30.81 & 49.02 & 26.57 & 47.79 & 33.30 & 57.06 & 30.23 & 51.29 \\
        CIReVL + SoFT (0.9) & 31.06 & 50.54 & 27.32 & 48.39 & 34.27 & 57.83 & 30.88 & 52.23 \\
        CIReVL + SoFT (1.0) & \textbf{31.26} & \textbf{50.98} & \textbf{27.52} & \textbf{49.13} & \textbf{34.37} & \textbf{58.44} & \textbf{31.05} & \textbf{52.85} \\
        \midrule
        \multicolumn{9}{l}{\textbf{ViT-L/14}} \\
        SEARLE & 26.89 & \textbf{45.58} & 20.48 & 43.13 & \textbf{29.32} & 49.97 & \textbf{25.56} & 46.23 \\
        SEARLE + SoFT (0.1) & \textbf{27.13} & 44.85 & \textbf{20.87} & \textbf{44.17} & 28.25 & \textbf{51.30} & 25.42 & \textbf{46.77} \\
        SEARLE + SoFT (0.2) & 26.30 & 43.67 & 20.77 & 42.94 & 27.94 & 49.97 & 25.00 & 45.53 \\
        SEARLE + SoFT (0.3) & 25.07 & 42.59 & 19.93 & 41.25 & 26.82 & 48.60 & 23.94 & 44.15 \\
        SEARLE + SoFT (0.5) & 23.21 & 39.40 & 18.49 & 39.07 & 24.63 & 45.39 & 22.11 & 41.29 \\
        SEARLE + SoFT (0.7) & 21.74 & 37.10 & 17.60 & 36.64 & 22.69 & 43.04 & 20.68 & 38.93 \\
        SEARLE + SoFT (0.9) & 20.56 & 35.77 & 17.20 & 35.35 & 21.21 & 40.59 & 19.66 & 37.24 \\
        CIReVL & 29.49 & 47.40 & 24.79 & 44.76 & 31.36 & 53.65 & 28.55 & 48.57 \\
        CIReVL + SoFT (0.1) & 30.03 & 48.14 & 24.59 & 44.47 & 31.36 & 54.31 & 28.66 & 48.97 \\
        CIReVL + SoFT (0.3) & 30.72 & 48.72 & 25.43 & 44.67 & 31.97 & 54.82 & 29.37 & 49.40 \\
        CIReVL + SoFT (0.5) & 31.21 & 49.46 & 25.88 & 45.41 & 32.69 & 55.69 & 29.93 & 50.19 \\
        CIReVL + SoFT (0.7) & 31.84 & 50.25 & 26.28 & 46.36 & 34.01 & 56.45 & 30.71 & 51.02 \\
        CIReVL + SoFT (0.9) & 32.04 & 51.37 & 27.37 & 47.30 & 34.73 & 57.37 & 31.38 & 52.01 \\
        CIReVL + SoFT (1.0) & \textbf{32.34} & \textbf{51.62} & \textbf{27.51} & \textbf{47.79} & \textbf{35.19} & \textbf{58.18} & \textbf{31.68} & \textbf{52.53} \\
        \bottomrule
    \end{tabular}
    }
    \caption{Effect of interpolation weight $\lambda$ on FashionIQ retrieval performance.}
    \label{tab:lambda-fiq}
\end{table*}

\begin{table*}
    \centering
    {
        \begin{tabular}{ll|cccc}
        \toprule
        \multicolumn{2}{c|}{Benchmark} & \multicolumn{4}{c}{\textbf{CIRCO (mAP@k)}} \\
        Backbone & Method & @5 & @10 & @25 & @50 \\
        \midrule
        \multirow{8}{*}{ViT-B/32} 
        & SEARLE & 9.35 & 9.94 & 11.13 & 11.84 \\
        & SEARLE + Penalty & 0.53 & 0.58 & 0.62 & 0.65 \\
        & SEARLE + Reward & 13.36 & 13.92 & 15.36 & 16.15 \\
        & SEARLE + SoFT & 12.57 & 13.12 & 14.38 & 15.17 \\
        & CIReVL & 14.94 & 15.42 & 17.00 & 17.82 \\
        & CIReVL + Penalty & 17.62 & 18.28 & 19.93 & 20.83 \\
        & CIReVL + Reward & 17.79 & 18.47 & 20.15 & 21.04 \\
        & CIReVL + SoFT & 19.21 & 20.04 & 21.86 & 22.75 \\
        \midrule
        \multirow{8}{*}{ViT-L/14} 
        & SEARLE & 11.68 & 12.73 & 14.33 & 15.12 \\
        & SEARLE + Penalty & 0.28 & 0.27 & 0.28 & 0.28 \\
        & SEARLE + Reward & 18.06 & 19.10 & 21.17 & 22.15 \\
        & SEARLE + SoFT & 15.72 & 16.45 & 18.06 & 18.93 \\
        & CIReVL & 18.57 & 19.01 & 20.89 & 21.80 \\
        & CIReVL + Penalty & 21.83 & 22.37 & 24.58 & 25.57 \\
        & CIReVL + Reward & 22.19 & 22.86 & 25.01 & 26.00 \\
        & CIReVL + SoFT & 23.90 & 24.72 & 26.94 & 27.93 \\
        \bottomrule
        \end{tabular}
    }
    \caption{Component-wise analysis of SoFT on the CIRCO benchmark (mAP@k).}
    \label{tab:component-circo}
\end{table*}

\begin{table*}
    \centering
    {\setlength{\tabcolsep}{1.8mm}
    \begin{tabular}{ll|cccc|ccc}
        \toprule
        \multicolumn{2}{c|}{Benchmark} & \multicolumn{4}{c|}{\textbf{CIRR}} & \multicolumn{3}{c}{\textbf{CIRR Subset}} \\
        Backbone & Method & R@1 & R@5 & R@10 & R@50 & R@1 & R@2 & R@3 \\
        \midrule
        \multirow{8}{*}{ViT-B/32}
        & SEARLE & 24.00 & 53.42 & 66.82 & 89.78 & 54.89 & 76.60 & 88.19 \\
        & SEARLE + Penalty & 7.95 & 19.10 & 26.19 & 46.41 & 45.52 & 65.21 & 79.21 \\
        & SEARLE + Reward & 27.45 & 58.87 & 72.10 & 91.85 & 60.48 & 80.29 & 91.04 \\
        & SEARLE + SoFT & 28.05 & 58.60 & 71.88 & 91.78 & 64.29 & 82.19 & 91.61 \\
        & CIReVL & 23.94 & 52.51 & 66.00 & 86.95 & 60.17 & 80.05 & 90.19 \\
        & CIReVL + Penalty & 30.41 & 60.96 & 72.05 & 90.22 & 65.93 & 83.78 & 92.46 \\
        & CIReVL + Reward & 30.94 & 61.21 & 72.53 & 90.46 & 66.00 & 83.78 & 92.41 \\
        & CIReVL + SoFT & 32.94 & 62.92 & 74.17 & 90.70 & 70.31 & 86.58 & 93.78 \\
        \midrule
        \multirow{8}{*}{ViT-L/14}
        & SEARLE & 24.24 & 52.48 & 66.29 & 88.84 & 53.76 & 75.01 & 88.19 \\
        & SEARLE + Penalty & 4.82 & 11.25 & 15.49 & 27.28 & 36.36 & 54.36 & 69.57 \\
        & SEARLE + Reward & 29.08 & 58.75 & 70.84 & 90.92 & 61.66 & 80.70 & 90.48 \\
        & SEARLE + SoFT & 30.29 & 59.74 & 71.49 & 90.65 & 65.47 & 83.23 & 92.29 \\
        & CIReVL & 24.55 & 52.31 & 64.92 & 86.34 & 59.54 & 79.88 & 89.69 \\
        & CIReVL + Penalty & 32.92 & 63.11 & 74.51 & 91.42 & 67.40 & 86.00 & 93.49 \\
        & CIReVL + Reward & 32.80 & 63.45 & 74.99 & 91.59 & 67.33 & 85.93 & 93.45 \\
        & CIReVL + SoFT & 35.54 & 65.25 & 76.41 & 91.95 & 71.59 & 87.64 & 94.15 \\
        \bottomrule
    \end{tabular}
    }
    \caption{Component-wise analysis of SoFT on the CIRR benchmark (Recall@k).}
    \label{tab:component-cirr}
\end{table*}

\begin{table*}
    \centering
    {\setlength{\tabcolsep}{2mm}
    \begin{tabular}{ll|cc|cc|cc|cc}
        \toprule
        \multicolumn{2}{c|}{Benchmark} & \multicolumn{2}{c|}{\textbf{Shirt}} & \multicolumn{2}{c|}{\textbf{Dress}} & \multicolumn{2}{c|}{\textbf{TopTee}} & \multicolumn{2}{c}{\textbf{Average}} \\
        Backbone & Method & R@10 & R@50 & R@10 & R@50 & R@10 & R@50 & R@10 & R@50 \\
        \midrule
        \multirow{8}{*}{ViT-B/32}
        & SEARLE & 24.44 & 41.61 & 18.54 & 39.51 & 25.70 & 46.46 & 22.89 & 42.53 \\
        & SEARLE + Penalty & 5.79 & 11.83 & 5.35 & 14.63 & 7.09 & 14.94 & 6.08 & 13.80 \\
        & SEARLE + Reward & 24.09 & 40.87 & 19.83 & 39.86 & 25.50 & 46.86 & 23.14 & 42.53 \\
        & SEARLE + SoFT & 24.29 & 42.64 & 19.48 & 40.46 & 25.54 & 47.78 & 23.10 & 43.63 \\
        & CIReVL & 28.36 & 47.84 & 25.29 & 46.36 & 31.21 & 53.85 & 28.29 & 49.35 \\
        & CIReVL + Penalty & 28.31 & 46.36 & 24.89 & 45.71 & 30.14 & 53.49 & 27.78 & 48.52 \\
        & CIReVL + Reward & 32.48 & 53.82 & 28.81 & 51.41 & 37.84 & 61.70 & 33.04 & 55.64 \\
        & CIReVL + SoFT & 31.26 & 50.98 & 27.52 & 49.13 & 34.37 & 58.44 & 31.05 & 52.85 \\
        \midrule
        \multirow{8}{*}{ViT-L/14}
        & SEARLE & 26.89 & 45.58 & 20.48 & 43.13 & 29.32 & 49.97 & 25.56 & 46.23 \\
        & SEARLE + Penalty & 3.97 & 7.75 & 2.63 & 6.74 & 4.18 & 8.41 & 3.59 & 7.63 \\
        & SEARLE + Reward & 27.38 & 44.11 & 20.33 & 43.58 & 29.07 & 51.56 & 25.59 & 46.42 \\
        & SEARLE + SoFT & 26.30 & 43.67 & 20.77 & 42.94 & 27.94 & 49.97 & 25.00 & 45.53 \\
        & CIReVL & 29.49 & 47.40 & 24.79 & 44.76 & 31.36 & 53.65 & 28.55 & 48.57 \\
        & CIReVL + Penalty & 30.42 & 48.33 & 24.54 & 45.86 & 31.36 & 53.19 & 28.77 & 47.13 \\
        & CIReVL + Reward & 32.92 & 51.67 & 27.41 & 48.49 & 37.38 & 60.22 & 32.57 & 53.46 \\
        & CIReVL + SoFT & 32.34 & 51.62 & 27.51 & 47.79 & 35.19 & 58.18 & 31.68 & 52.53 \\
        \bottomrule
    \end{tabular}
    }
    \caption{Component-wise analysis of SoFT on the FashionIQ benchmark (Recall@k).}
    \label{tab:component-fiq}
\end{table*}

\begin{table*}[t]
    \centering
    \setlength{\tabcolsep}{3pt}
    \renewcommand{\arraystretch}{1.2}
    \begin{tabular}{ll|cccc|cccc|ccc}
    \toprule
    \multicolumn{2}{c|}{\textbf{Benchmark}} & \multicolumn{4}{c|}{\textbf{CIRCO (mAP@K)}} & \multicolumn{4}{c|}{\textbf{CIRR (Recall@K)}} & \multicolumn{3}{c}{\textbf{CIRR (Recall$_\text{subset}$@K)}} \\
    \textbf{Backbone} & \textbf{Method} & k=5 & k=10 & k=25 & k=50 & k=1 & k=5 & k=10 & k=50 & k=1 & k=2 & k=3 \\
    \midrule
    \multirow{4}{*}{ViT-B/32}
    & CIReVL & 14.94 & 15.42 & 17.00 & 17.82 & 23.94 & 52.51 & 66.00 & 86.95 & 60.17 & 80.05 & 90.19 \\
    & {CIReVL + SoFT} & {19.21} & {20.04} & {21.86} & {22.75} & {32.94} & {62.92} & {74.17} & {90.70} & {70.31} & {86.58} & {93.78} \\
    & CIReVL(4o) & 19.14	&20.01&	21.92&	22.76 & 32.51&	63.18	&73.81	&91.88	&67.42	&85.23	&93.42 \\
    & {CIReVL(4o) + SoFT} & 20.80&	21.63	&23.54	&24.49 & 34.05	&64.72	&75.35&	91.57&	71.18&	87.25	&94.41 \\
    
    \midrule
    \multirow{4}{*}{ViT-L/14}
    & CIReVL & 18.57 & 19.01 & 20.89 & 21.80 & 24.55 & 52.31 & 64.92 & 86.34 & 59.54 & 79.88 & 89.69 \\
    & {CIReVL + SoFT} & {23.90} & 24.72 & 26.94 & 27.93 & {35.54} & {65.25} & {76.41} & {91.95} & {71.59} & {87.64} & {94.15} \\
    & CIReVL(4o) & 23.88&	24.54	&26.85&	27.93 & 34.94&	64.29	&75.88&	92.68	&69.33	&86.60&	94.24 \\
    & {CIReVL(4o) + SoFT} & 25.57&	26.36&	28.82	&29.81 & 36.70	&65.93&	77.54	&92.75	&72.96&	88.60	&94.89 \\
    
    \bottomrule
    \end{tabular}
    \caption{Quantitative results on CIRCO and CIRR under a unified LLM setting (both CIReVL and SoFT using GPT-4o).}
    \label{tab:unified-llm-circo-cirr}
\end{table*}

\begin{table*}[t]
    \centering
    \setlength{\tabcolsep}{3pt}
    \begin{tabular}{ll|cc|cc|cc|cc}
    \toprule
    \multicolumn{2}{c|}{\textbf{Benchmark}} & \multicolumn{2}{c|}{\textbf{Shirt}} & \multicolumn{2}{c|}{\textbf{Dress}} & \multicolumn{2}{c|}{\textbf{TopTee}} & \multicolumn{2}{c}{\textbf{Average}} \\
    \textbf{Backbone} & \textbf{Method} & R@10 & R@50 & R@10 & R@50 & R@10 & R@50 & R@10 & R@50 \\
    \midrule
    \multirow{4}{*}{ViT-B/32}
    & CIReVL & 28.36&	47.84	&25.29&	46.36	&31.21	&53.85&	28.29	&49.35 \\
    & CIReVL + SoFT &31.26&	50.98	&27.52&	49.13	&34.37	&58.44&	31.05
	&52.85 \\
    & CIReVL(4o) & 28.80	&47.25	&25.48&	47.35	&31.82&	54.11	&28.70	&49.57 \\
    & {CIReVL(4o) + SoFT} &30.62&	50.69	&28.51	&49.63&	34.52	&57.78	&31.22&	52.70 \\
    
    \midrule
    \multirow{4}{*}{ViT-L/14}
    & CIReVL & 29.49&	47.40	&24.79	&44.76&	31.36&	53.65	&28.55	&48.57 \\
    & CIReVL + SoFT & 32.34&	51.62	&27.51 &	47.79	&35.19	&58.18	&31.68	&52.53 \\
    & CIReVL(4o) & 29.54 & 46.81 &25.53	& 45.36	&31.26 & 54.51	& 28.78 & 48.89 \\
    & {CIReVL(4o) + SoFT} & 32.43	& 50.88 & 28.06	& 48.54 & 35.08	& 58.18 & 31.86 &	52.53 \\
    \bottomrule
    \end{tabular}
    \caption{Quantitative results on FashionIQ under a unified LLM setting (both CIReVL and SoFT using GPT-4o).}
    \label{tab:unified-llm-fiq}
\end{table*}

\subsection{Unified Large Language Model Consistency Check}
To ensure that the observed improvement does not arise from differences in the underlying language model, we unified both CIReVL and SoFT to use GPT-4o as their textual reasoning component.
Table~\ref{tab:unified-llm-circo-cirr} and Table~\ref{tab:unified-llm-fiq} show that SoFT consistently improves retrieval performance 
across all benchmarks, confirming that the gain stems from its filtering mechanism rather than from the capability gap between Large Language Models(LLMs).

\section{Multi-Target Triplet Dataset Pipeline}

\subsection{Large Language Model Prompt Templates}
\subsubsection{Prompt Templates for Query Generation.}
To comprehensively capture plausible multi-target candidates during candidate retrieval, we design prompt templates that produce two complementary queries from each input pair(reference image, modification text). Rather than serving as simple reformulations, these templates represent distinct semantic perspectives of user intent, helping to prevent the omission of valid yet diverse candidate targets.

\begin{itemize}
\item \textbf{Text-focused query}: Captures only the intended modification described in the text.
\item \textbf{Compositional query}: Integrates the modification with compatible attributes inferred from the reference image.
\end{itemize}

Distinct templates are used for CIRR and FashionIQ to account for domain-specific characteristics and annotation styles.

\subsubsection{Prompt Templates for Confidence Scoring of Candidate Groups.} 
For each (reference image, modification text) pair, we form three candidate groups based on distinct retrieval criteria: \textit{Textual to Modification}, \textit{Compositional}, and \textit{Visual Similarity}.  
Each group is then evaluated by an LLM using a unified prompt structure that receives the reference image, the modification text, and the candidate images within the group as input.  
The LLM assigns a confidence score between 0.0 and 1.0 to each image, indicating how well it aligns with the intended modification.

\subsubsection{Prompt Template for Refining Single-Target Modification Text.}
To support precise evaluation under reduced ambiguity, we refine the original modification text so that it describes a single target image explicitly. This process ensures that the refined text is not only faithful to the original intent but also discriminative against plausible distractors.

\begin{tcolorbox}[colback=blue!5,colframe=blue!40!black,title=Query Generation Prompt Template - CIRR]
You are given two pieces of input: \\
1. A reference image description that tells you what the original image looks like. \\
2. A User's Modifications text that describes how the user wants the image to be changed. \\

Your task is to write a two-sentence description of the image the user wants.

- The sentence1 should focus on the only User's modifications—describe the image primarily according to the User's Modifications. \\
- The sentence2 should preserve the elements from the reference image that do **not** conflict with the User's Modifications, describing details from the original image that can be retained. \\
  If there are no additional elements to preserve beyond what is stated in the User's Modifications, return an empty string. \\
  
Describe the image using only concrete, observable attributes.
Make sure both sentences are concise and consistent. \\

\#\# User's Modifications \\
``\{caption1\}'' \\
\end{tcolorbox}

\begin{tcolorbox}[colback=blue!5,colframe=blue!40!black,title=Query Generation Prompt Template - FashionIQ]
You are given two pieces of input: \\
1. A reference image description that tells you what the original product looks like. \\
2. A User's Modifications text that describes how the user wants the product to be changed. \\

Your task is to write a two-sentence description of the product the user wants.

- The sentence1 should focus on the only User's modifications—describe the product primarily according to the User's Modifications. \\
- The sentence2 should preserve the elements from the reference image that do **not** conflict with the User's Modifications, describing details from the original product that can be retained. \\
  
Describe the image using only concrete, observable attributes (e.g., color, shape, texture, pattern, material).
Avoid indirect or relative terms like ``more,'' ``less'', ``similar'', or ``retain''. Use precise, objective language.
Make sure both sentences are concise and consistent.\\

\#\# User's Modifications \\
1. ``\{caption1\}'' \\
2: ``\{caption2\}''
\end{tcolorbox}

\begin{tcolorbox}[colback=green!5,colframe=green!40!black,title=Confidence Scoring Prompt Template]
You are an AI assistant specialized in image analysis and candidate selection. Your task is to analyze images and select the most appropriate candidates based on given criteria. \\

TASK INPUT: \\
- reference\_image\_name: \{ref\_image\_name\} \\
- relative\_captions: \{relative\_captions\} \\
- candidate\_images: \{top\_k\_names\} \\

TASK INSTRUCTIONS: \\
1. Analyze the reference image to understand its key visual features and context. \\
2. Interpret the relative captions carefully — they describe how the desired candidate image should differ from or relate to the reference image. \\
3. For each candidate image: \\
    - Evaluate how well it visually represents the *expected relationship* or *transformation* implied by combining the reference image and the relative captions. \\
    - Assign a confidence score between 0.0 and 1.0, reflecting how accurately the candidate image captures this intended change or relation. \\
\end{tcolorbox}

\begin{tcolorbox}[colback=yellow!5,colframe=yellow!40!black,title=Modification Text Refinement Prompt Template]
You are an expert caption-writer for composed-image retrieval across various domains. \\

The images above show: \\
1. Reference image \\
2. Target image \\
3. Comparison images (similar but incorrect) \\

Original captions: \{original\_captions\} \\
The original captions describe how to transform the reference image into the target image, but they are not specific enough to rule out the comparison images. \\

Task: Write one refined caption that uniquely identifies the target image while staying faithful to the intent of the original captions. \\

Guidelines: \\
- Use the original captions as a foundation; keep their meaning while making them more specific. \\
- Preserve the original captions' writing tone and sentence style. \\
- Add concrete, observable details present in the target but absent from the comparison images. \\
- Be concise and output exactly one sentence. \\
\end{tcolorbox}

\subsection{Quantitative Results}
\subsubsection{Multi-Target and CIRR and FashionIQ.}
We evaluate SoFT on the Multi-Target versions of two standard CIR benchmarks: CIRR and FashionIQ. Experiments are conducted using two baseline retrieval models—CIReVL and SEARLE—with and without the proposed SoFT module. We vary the interpolation weight \(\lambda \in \{0.1, 0.3, 0.5, 0.7, 0.9, 1.0\}\), which adjusts the relative contribution of the constraint-based reweighting to the base model score. Values in parentheses in the tables denote the \(\lambda\) used. Performance is reported in terms of R@5, R@10, R@25, and R@50.

\subsubsection{Impact of Target Annotation on SoFT Effectiveness.}
We observe a clear difference in the stability of SoFT’s effectiveness between the standard and Multi-Target versions of FashionIQ, particularly when SEARLE is used as the base retriever. As shown in Table~\ref{tab:lambda-fiq}, under the original single-target benchmark, applying SoFT does not consistently improve performance: in several cases, the baseline retriever without SoFT achieves the highest score. In contrast, results reported in Tables~\ref{tab:mt-cirr} through~\ref{tab:mt-fiq-avg} show that under the Multi-Target setting, SoFT consistently improves retrieval performance across categories and evaluation metrics. This trend holds regardless of the specific value of $\lambda$ or the underlying retriever, indicating that SoFT’s effectiveness becomes substantially more reliable when multiple valid targets are taken into account.

This discrepancy is attributed to differences in target annotation scope. In the original FashionIQ benchmark, each query is paired with a single annotated target, despite the existence of multiple semantically valid candidates. Consequently, even when SoFT promotes images that better satisfy the modification intent, such retrievals may not be reflected as correct during evaluation, leading to apparent instability or diminished gains. By explicitly incorporating multiple valid targets per query, the Multi-Target benchmark alleviates this issue and provides a more faithful assessment of SoFT’s impact.

Overall, these results suggest that the observed instability of SoFT in the standard FashionIQ setting stems largely from annotation underspecification, rather than from the filtering mechanism itself. When evaluation protocols better align with the open-ended nature of user intent, SoFT demonstrates consistent and robust improvements across a wide range of configurations.

\subsection{Qualitative Results}
\begin{figure*}[t]
    \centering    \includegraphics[width=\linewidth]{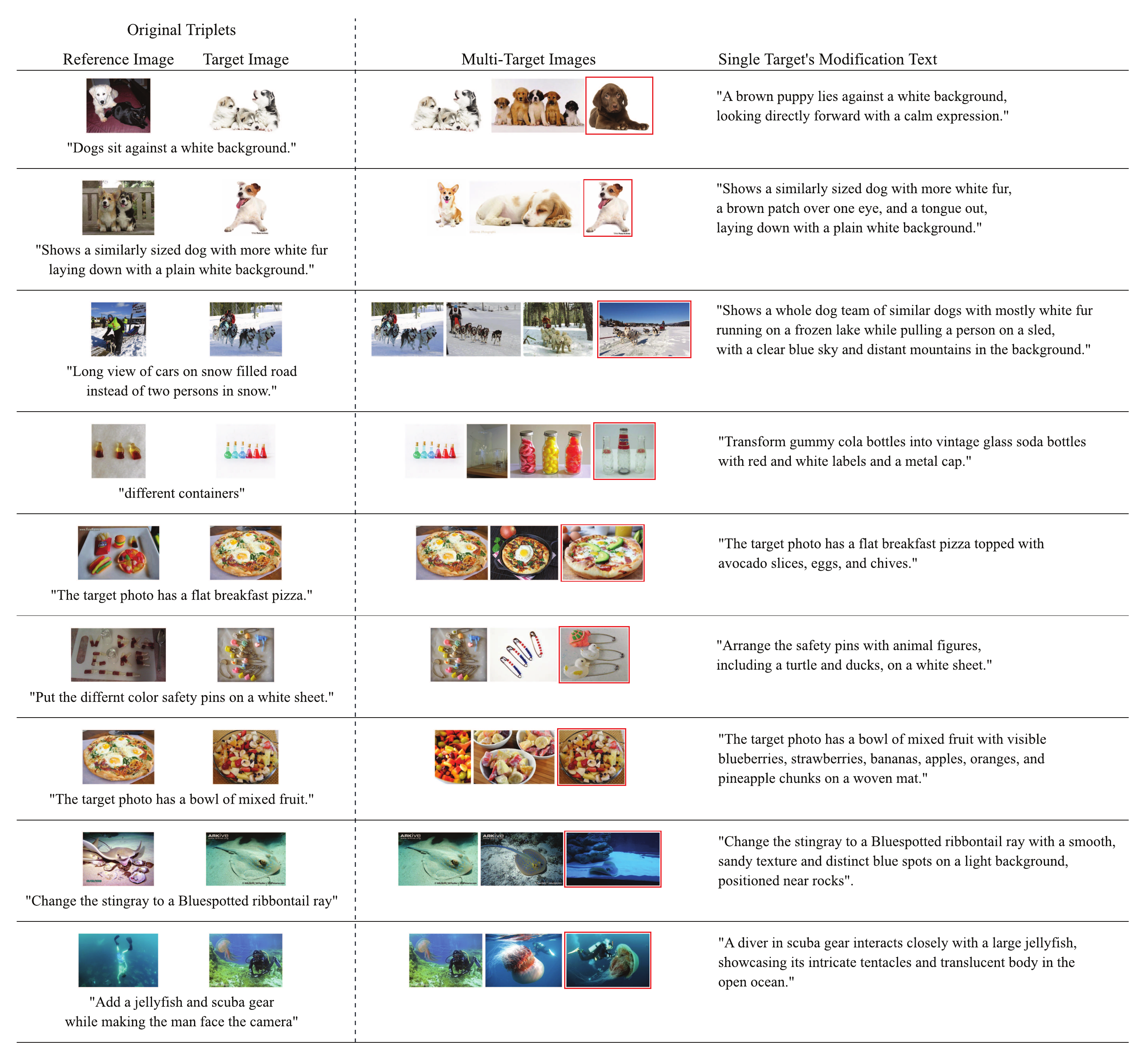}
    \caption{
        Examples from the multi-target CIRR dataset. The original triplet is shown on the left, followed by multi-target results and a refined text for the selected target (red box).
    }
    \label{fig:mt_cirr_examples}
\end{figure*}

\begin{figure*}[t]
    \centering    \includegraphics[width=\linewidth]{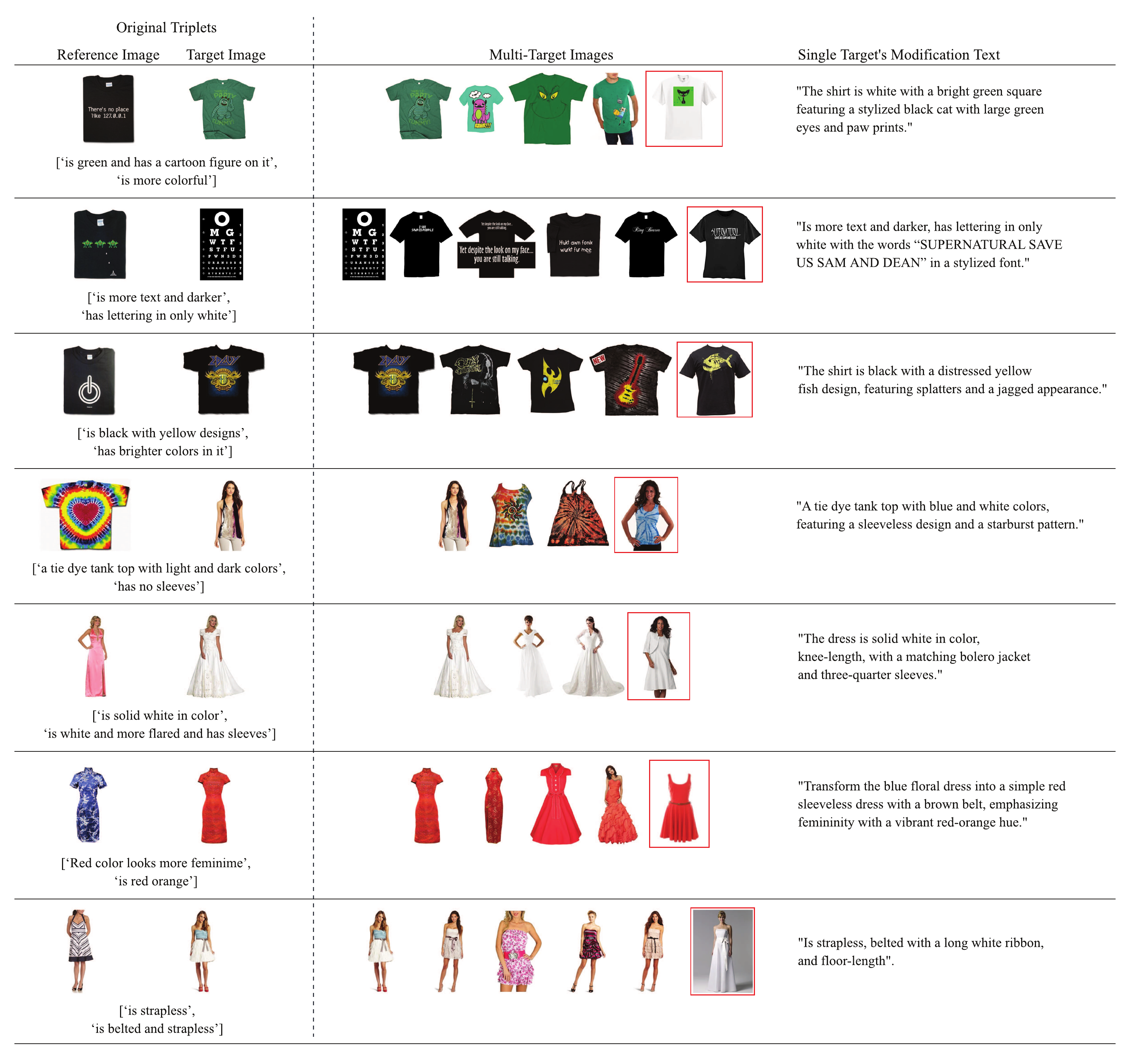}
    \caption{
        Examples from the multi-target FashionIQ dataset. The original triplet is shown on the left, followed by multi-target results and a refined text for the selected target (red box).
    }
    \label{fig:mt_fiq_examples}
\end{figure*}

We present examples from our multi-target versions of CIRR and FashionIQ in Figure~\ref{fig:mt_cirr_examples} and~\ref{fig:mt_fiq_examples}. Each example includes the original triplet, a set of valid target images, and a refined modification text for a randomly selected target (highlighted in red box). These examples illustrate the ambiguity in user intent and how our pipeline resolves it into precise single-target queries.

\begin{table}
    \centering
    \begin{tabular}{l|cccc}
    \toprule
     Method & \multicolumn{4}{c}{\textbf{CIRR (mAP@k)}} \\
    ViT-L/14 & @5 & @10 & @25 & @50 \\
    \midrule
    CIReVL & 58.04 & 56.31 & 53.8 &	52.33 \\
    CIReVL + SoFT (0.1) & 58.31	& 56.54	& 53.91	& 52.53\\
    CIReVL + SoFT (0.3) & 58.75	& 56.93	& 54.31	& 52.93 \\
    CIReVL + SoFT (0.5) & 59.53	& 57.66	& 54.94	& 53.53 \\
    CIReVL + SoFT (0.7) & 60.21	& 58.23	& 55.47	& 54.01 \\
    CIReVL + SoFT (0.9) & 60.82	& 58.77	& 55.92	& 54.45 \\
    CIReVL + SoFT (1.0) & 61.12	& 59.08	& 56.24	& 54.74 \\
    SEARLE & 52.10 & 50.86 & 48.15 & 46.70 \\
    SEARLE + SoFT (0.1) & 60.22 & 58.13 & 54.81 & 53.38 \\
    SEARLE + SoFT (0.3) & 62.29 & 59.66 & 56.60 & 54.99 \\
    SEARLE + SoFT (0.5) & 59.13 & 56.90 & 53.86 & 52.17 \\
    SEARLE + SoFT (0.7) & 55.35 & 53.64 & 50.81 & 48.94 \\
    SEARLE + SoFT (0.9) & 52.02 & 50.45 & 47.74 & 45.99 \\
    SEARLE + SoFT (1.0) & 50.35 & 49.02 & 46.40 & 44.60 \\    
    \bottomrule
    \end{tabular}
    
    \caption{Multi-Target Evaluation on CIRR.}
    \label{tab:mt-cirr}
\end{table}

\begin{table}
    \centering
    {\setlength{\tabcolsep}{1.8mm}
        \begin{tabular}{l|cccc}
        \toprule
         Method & \multicolumn{4}{c}{\textbf{Shirt (mAP@k)}} \\
        ViT-L/14 & @5 & @10 & @25 & @50 \\
        \midrule
        CIReVL & 24.14 & 24.31 & 23.36 & 22.01 \\
        CIReVL + SoFT (0.1) & 24.44 & 24.59 & 23.58 &22.23 \\
        CIReVL + SoFT (0.3) & 25.31 & 25.39 & 24.29 & 22.81 \\
        CIReVL + SoFT (0.5) & 26.00 & 26.02 & 24.85 & 23.21 \\
        CIReVL + SoFT (0.7) & 26.76 & 26.73 & 25.53 & 23.93 \\
        CIReVL + SoFT (0.9) & 27.72 & 27.77 & 26.25 & 24.67 \\
        CIReVL + SoFT (1.0) & 28.17	& 28.15 & 26.67 & 25.05 \\
        SEARLE & 41.64 & 40.03 & 36.54 & 33.61 \\
        SEARLE + SoFT (0.1) & 42.07	& 40.39 & 36.88 & 33.95 \\
        SEARLE + SoFT (0.3) & 43.21	& 41.71 & 37.76 & 35.01 \\
        SEARLE + SoFT (0.5) & 44.49	& 42.78 & 38.67 & 35.91 \\
        SEARLE + SoFT (0.7) & 45.52 & 43.65	& 39.61	& 36.72 \\
        SEARLE + SoFT (0.9) & 46.55	& 44.58	& 40.34	& 37.53 \\
        SEARLE + SoFT (1.0) & 46.82	& 44.85 & 40.51 & 37.68 \\
        \bottomrule
        \end{tabular}
    }
    \caption{Multi-Target Evaluation on FashionIQ - Shirt.}
    \label{tab:mt-fiq-shirt}
\end{table}

\begin{table}
    \centering
    {\setlength{\tabcolsep}{1.8mm}
        \begin{tabular}{l|cccc}
        \toprule
         Method & \multicolumn{4}{c}{\textbf{Dress (mAP@k)}} \\
        ViT-L/14 & @5 & @10 & @25 & @50 \\
        \midrule
        CIReVL & 20.71 & 21.17 & 19.62 & 17.90 \\
        CIReVL + SoFT (0.1) & 21.02	& 21.41	& 19.78	& 18.10 \\
        CIReVL + SoFT (0.3) & 21.65	& 21.75	& 20.27 & 18.48 \\
        CIReVL + SoFT (0.5) & 22.43	& 22.35	& 20.70	& 18.89 \\
        CIReVL + SoFT (0.7) & 23.16	& 23.12	& 21.42	& 19.47 \\
        CIReVL + SoFT (0.9) & 23.83	& 23.70	& 22.05	& 19.97 \\
        CIReVL + SoFT (1.0) & 24.39	& 24.18	& 22.31	& 20.30 \\
        SEARLE & 36.26 & 35.42 & 31.46 & 28.21 \\
        SEARLE + SoFT (0.1) & 36.96	& 35.92	& 31.96	& 28.64 \\
        SEARLE + SoFT (0.3) & 37.89	& 36.65	& 32.71	& 29.33 \\
        SEARLE + SoFT (0.5) & 38.90	& 37.63	& 33.40	& 30.05 \\
        SEARLE + SoFT (0.7) & 39.78	& 38.33	& 34.04	& 30.80 \\
        SEARLE + SoFT (0.9) & 40.31	& 38.67	& 34.47	& 30.98 \\
        SEARLE + SoFT (1.0) & 40.54	& 38.91	& 34.80	& 31.22 \\
        \bottomrule
        \end{tabular}
    }
    \caption{Multi-Target Evaluation on FashionIQ - Dress.}
    \label{tab:mt-fiq-dress}
\end{table}

\begin{table}
    \centering
    {\setlength{\tabcolsep}{1.8mm}
        \begin{tabular}{l|cccc}
        \toprule
         Method & \multicolumn{4}{c}{\textbf{Toptee (mAP@k)}} \\
        ViT-L/14 & @5 & @10 & @25 & @50 \\
        \midrule
        CIReVL & 23.85 & 24.29 & 23.17	& 21.63 \\
        CIReVL + SoFT (0.1) & 24.20	& 24.69	& 23.48	& 21.92 \\
        CIReVL + SoFT (0.3) & 25.25	& 25.53	& 24.29	& 22.52 \\
        CIReVL + SoFT (0.5) & 26.05	& 26.17	& 24.83	& 23.10 \\
        CIReVL + SoFT (0.7) & 27.05	& 27.13	& 25.62	& 23.82 \\
        CIReVL + SoFT (0.9) & 28.17	& 28.21	& 26.52	& 24.62 \\
        CIReVL + SoFT (1.0) & 28.86	& 28.78	& 27.05	& 25.13 \\
        SEARLE & 44.44	& 42.62	& 38.53	& 35.20 \\
        SEARLE + SoFT (0.1) & 44.86	& 42.96	& 38.73	& 35.54 \\
        SEARLE + SoFT (0.3) & 45.92	& 43.80	& 39.51	& 36.12 \\
        SEARLE + SoFT (0.5) & 46.89	& 44.86	& 40.34	& 36.92 \\
        SEARLE + SoFT (0.7) & 47.90	& 45.83	& 41.24	& 37.71 \\
        SEARLE + SoFT (0.9) & 48.80 & 46.51 & 41.66	& 38.04 \\
        SEARLE + SoFT (1.0) & 49.15	& 46.84	& 41.85	& 38.19 \\
        \bottomrule
        \end{tabular}
    }
    \caption{Multi-Target Evaluation on FashionIQ - Toptee.}
    \label{tab:mt-fiq-toptee}
\end{table}

\begin{table}
    \centering
    \begin{tabular}{l|cccc}
    \toprule
     Method & \multicolumn{4}{c}{\textbf{Average (mAP@k)}} \\
    ViT-L/14 & @5 & @10 & @25 & @50 \\
    \midrule
    CIReVL & 22.90 & 23.26 & 22.05 & 20.51 \\
    CIReVL + SoFT (0.1) & 23.22	& 23.56	& 22.28	& 20.75 \\
    CIReVL + SoFT (0.3) & 24.07	& 24.22	& 22.95	& 21.27 \\
    CIReVL + SoFT (0.5) & 24.83	& 24.85	& 23.46	& 21.73 \\
    CIReVL + SoFT (0.7) & 25.66	& 25.66	& 24.19	& 22.41 \\
    CIReVL + SoFT (0.9) & 26.57	& 26.56	& 24.94	& 23.09 \\
    CIReVL + SoFT (1.0) & 27.14	& 27.04	& 25.34	& 23.49 \\
    SEARLE & 40.78   & 39.36  & 35.51  & 32.34  \\
    SEARLE + SoFT (0.1) &  41.30 & 39.76 & 35.86 & 32.71 \\
    SEARLE + SoFT (0.3) & 42.34	& 40.72	& 36.66	& 33.49 \\
    SEARLE + SoFT (0.5) & 43.43	& 41.76	& 37.47	& 34.29 \\
    SEARLE + SoFT (0.7) & 44.40	& 42.60	& 38.30	& 35.08 \\
    SEARLE + SoFT (0.9) & 45.22	& 43.25	& 38.82	& 35.52 \\
    SEARLE + SoFT (1.0) & 45.50	& 43.53	& 39.05	& 35.70 \\
    \bottomrule
    \end{tabular}
    \caption{Multi-Target Evaluation on FashionIQ - Average.}
    \label{tab:mt-fiq-avg}
\end{table}

% \begin{table}
%     \centering
%     {\setlength{\tabcolsep}{1.8mm}
%         \begin{tabular}{l|cccc}
%         \toprule
%          Method & \multicolumn{4}{c}{\textbf{Refined-CIRR (Recall@k)}} \\
%         ViT-L/14 & @1 & @5 & @10 & @50 \\
%         \midrule
%         CIReVL & 58.98 & 87.45 & 92.96 & 98.33 \\
%         CIReVL + SoFT (0.1) & 59.30 & 87.59 & 93.10 & 98.33 \\
%         CIReVL + SoFT (0.3) & 60.02 & 87.86 & 93.55 & 98.38 \\
%         CIReVL + SoFT (0.5) & 60.60 & 88.18 & 93.82 & 98.47 \\
%         CIReVL + SoFT (0.7) & 61.06 & 88.54 & 93.91 & 98.51 \\
%         CIReVL + SoFT (0.9) & 61.28 & 89.03 & 94.09 & 98.69 \\
%         CIReVL + SoFT (1.0) & 61.37 & 89.21 & 94.13 & 98.78 \\
%         SEARLE & 52.10 & 50.86 & 48.15 & 46.70 \\
%         SEARLE + SoFT (0.1) & 60.22 & 58.13 & 54.81 & 53.38 \\
%         SEARLE + SoFT (0.2) & 62.67 & 60.20 & 56.82 & 55.34 \\
%         SEARLE + SoFT (0.3) & 62.29 & 59.66 & 56.60 & 54.99 \\
%         SEARLE + SoFT (0.5) & 59.13 & 56.90 & 53.86 & 52.17 \\
%         SEARLE + SoFT (0.7) & 55.35 & 53.64 & 50.81 & 48.94 \\
%         SEARLE + SoFT (0.9) & 52.02 & 50.45 & 47.74 & 45.99 \\
%         SEARLE + SoFT (1.0) & 50.35 & 49.02 & 46.40 & 44.60 \\
%         \bottomrule
%         \end{tabular}
%     }
%     \caption{Multi-Target Evaluation on CIRR with mAP@k metrics.}
%     \label{tab:mt-cirr}
% \end{table}

\end{document}